\documentclass[twoside]{article}
\usepackage{qic,epsfig,amsmath}

\begin{document}
\setlength{\textheight}{8.0truein}    

\runninghead{Photon engineering for quantum information processing
$\ldots$} {A.B. U'Ren, K. Banaszek, I.A. Walmsley $\ldots$}

\normalsize\textlineskip
\thispagestyle{empty}
\setcounter{page}{1}

\date{May, 2003}

\alphfootnote

\fpage{1}

\centerline{\bf
PHOTON ENGINEERING FOR QUANTUM INFORMATION PROCESSING}
\vspace*{0.035truein} \centerline{\bf} \centerline{\footnotesize
A.B. U'REN\footnote{A.B. U'Ren's e-mail address is:
alfred.uren@physics.ox.ac.uk} , K. BANASZEK AND I.A. WALMSLEY}
\vspace*{0.015truein} \centerline{\footnotesize\it Clarendon
Laboratory, Oxford University, Oxford, OX1 3PU, UK}
\baselineskip=10pt \centerline{\footnotesize{May 2003}}
 \vspace*{0.225truein}


\abstracts{
We study distinguishing information in the context of quantum
interference involving more than one parametric downconversion
(PDC) source and in the context of polarization-entangled photon
pairs based on PDC. We arrive at specific design criteria for
two-photon sources so that when used as part of complex optical
systems, such as photon-based quantum information processing
schemes, distinguishing information between the photons is
eliminated guaranteeing high visibility interference.  We propose
practical techniques which lead to suitably engineered two-photon
states that can be realistically implemented with available
technology. Finally, we study an implementation of the
nonlinear-sign shift (NS) logic gate with PDC sources and show the
effect of distinguishing information on the performance of the
gate. }{}{}

 \vspace*{10pt}

 \vspace*{3pt}

\vspace*{1pt}\textlineskip    
\section{Introduction}           
\vspace*{-0.5pt} \noindent

Among physical systems that can be used for practical realizations
of novel information processing schemes based on intrinsically
quantum phenomena, photons are the primary candidates for
constituting carriers of quantum information. Applications of
photons as ``flying qubits'' include quantum cryptography,
communication between distributed computational nodes in quantum
networks, as well as elementary building blocks for registers in
all-optical quantum computation.

Using single photons as carriers of qubits in quantum information
processing requires, in most cases, that they be made to interact
in a controlled coherent way. This is usually facilitated by modal
interference, which is implemented in practice with the help of
passive linear optical elements such as beam splitters,
polarizers, and wave plates. Though the range of interactions that
can be realized by these means is restricted, the linear optical
approach is at present the most viable one, as nonlinear effects
are usually much too weak at the single photon level in
travelling-wave configurations. Modal interference is, however,
critically sensitive to the spatio-temporal structure of the
interfering photons. Great care must therefore be exercised in
making sure that no distinguishing information is contained in the
interfering photons that would allow one to trace their origin. A
common approach to cope with this problem is the implementation of
strong spatial and spectral filtering. Such a method, however,
reduces the available photon sample, and also contributes
deleteriously to the overall detection efficiency. Overcoming the
limitations of this approach is currently one of the main
challenges in the further development of quantum information
processing applications in the photonic domain.

The purpose of this paper is to explore methods for engineering
the modal structure of photon sources to ensure optimal
performance of quantum information applications. The physical
process that we will consider is spontaneous parametric
down-conversion which has been the primary source of nonclassical
optical radiation in recent experiments ranging from
implementations of quantum information technologies to fundamental
tests of quantum mechanics. Our goal is to develop photon sources
that produce photons in well specified single spatio-temporal
modes thus eliminating the usual need for filtering: the resulting
engineered photons would have identical mode structures, with in
principle no distinguishing information residing in any degree of
freedom, thus ensuring high visibility interference. We will base
our discussion on the analysis of well known interferometers, such
as the event-ready version of the Hong-Ou-Mandel interferometer
and the Braunstein-Mann Bell state analyzer. We will also analyze
the role of distinguishing information in the nonlinear sign-shift
gate that constitutes the basic element of the recently proposed
scheme for quantum computing based on linear
optics\cite{KLM,RalphPRA}.

\section{Criteria for the design of photon states}\label{Sec:criteria}

\subsection{Characterizing entanglement in photon
pairs.}\label{Sec:schmidt}

A two-photon PDC state can in general be expressed as a weighted
sum of creation operators acting on vacuum as follows:
\begin{equation}
\begin{split}
|\Psi\rangle = \sum_{\mu_s}\sum_{\mu_i} \int\limits_0^\infty
d\omega_s \int\limits_0^\infty
 d\omega_i
\int d{\bf k}_s^\bot \int d{\bf k}_i^\bot
S(\omega_s,{\bf k}_s^\bot,\mu_s ; \omega_i,{\bf k}_i^\bot,\mu_i)\\
\times\hat{a}^\dag_{{\bf k}^\bot_s \mu_s}
(\omega_s)\hat{a}^\dag_{{\bf k}^\bot_i
\mu_i}(\omega_i)|\mbox{vac}\rangle
\end{split} \label{E:state}
\end{equation}
where the function $S(\omega_s,{\bf k}_s^\bot,\mu_s ;
\omega_i,{\bf k}_i^\bot,\mu_i)$ represents the joint two-photon
amplitude in terms of the photon degrees of freedom which include:
frequency, transverse momentum (with respect to a fixed direction
of propagation defined by the pump field) and polarization. The
two-photon probability amplitude depends on the form of the
electromagnetic field used as a pump in the down-conversion
process, and on the so-called phase matching function, defined by
the optical properties of the crystal. A useful tool for analyzing
the entanglement present in the continuous degree of freedom is to
carry out a Schmidt decomposition \cite{cklaw} of the
spatio-temporal-polarization state amplitude $S({\bf k}^\bot_s
,\omega _s ,\mu _s ; {\bf k}^\bot_i ,\omega _i ,\mu _i )$ into a
sum of products of functions:
\begin{equation}
S( \omega _s ,{\bf k}^\bot_s,\mu_s;\omega_i,{\bf k}^\bot_i ,\mu_i)
= \sum_n {\sqrt {\lambda _n } \psi _n (} \omega _s,{\bf k}^\bot_s
,\mu_s )\phi _n (\omega _i,{\bf k}^\bot_i ,\mu_i)\label{E:schmidt}
\end{equation}
where $\omega_{s,i}$, ${\bf k}^\bot_{s,i}$ and $\mu_{s,i}$
represent the wavelength, transverse wavevector and polarization
of the signal (s) and idler (i) photons. The Schmidt functions
$\psi _n ( \omega _s,{\bf k}^\bot_s ,\mu_s )$ and $\phi _n (\omega
_i,{\bf k}^\bot_i ,\mu_i)$ can be thought of as the basic building
blocks of entanglement in the sense that if the signal photon is
determined to be described by a function $\psi_n$, we know with
certainty that its idler sibling is described by the corresponding
function $\phi_n$. The probability of finding this specific pair
of modes is given by the parameter $\lambda_n$, assumed to be real
and nonnegative. In general, obtaining such a Schmidt
decomposition entails solving certain integral equations
\cite{cklaw}. By approximating the sinc function in the phase
matching function as a Gaussian it becomes possible to carry out
the decomposition analytically in certain cases
\cite{WarrenAlfred} (see appendix A). It is found that the
resulting entanglement does not necessarily involve the infinitely
many degrees of freedom in the frequency-space-polarization
continuum with typically only a finite number of Schmidt modes
having appreciable eigenvalue magnitudes.  The amount of
entanglement can be conveniently quantified by the cooperativity
parameter \cite{eberlyschmidt} defined in terms of the Schmidt
eigenvalues as:
\begin{equation}
K=\frac{1}{\sum\limits_k \lambda_k^2}\label{E:K} \end{equation}

The value of $K$ gives an indication of the number of active
Schmidt mode pairs, which in turn is a measure of how much
entanglement (including spectral, spatial and polarization degrees
of freedom) is present in the photon pairs. A two-photon state for
which the cooperativity assumes its minimum allowed value $K=1$
represents a state in which there is a single pair of Schmidt
modes and therefore exhibits no spectral entanglement.

\begin{figure} [htbp]
\vspace*{13pt}
\centerline{\psfig{file=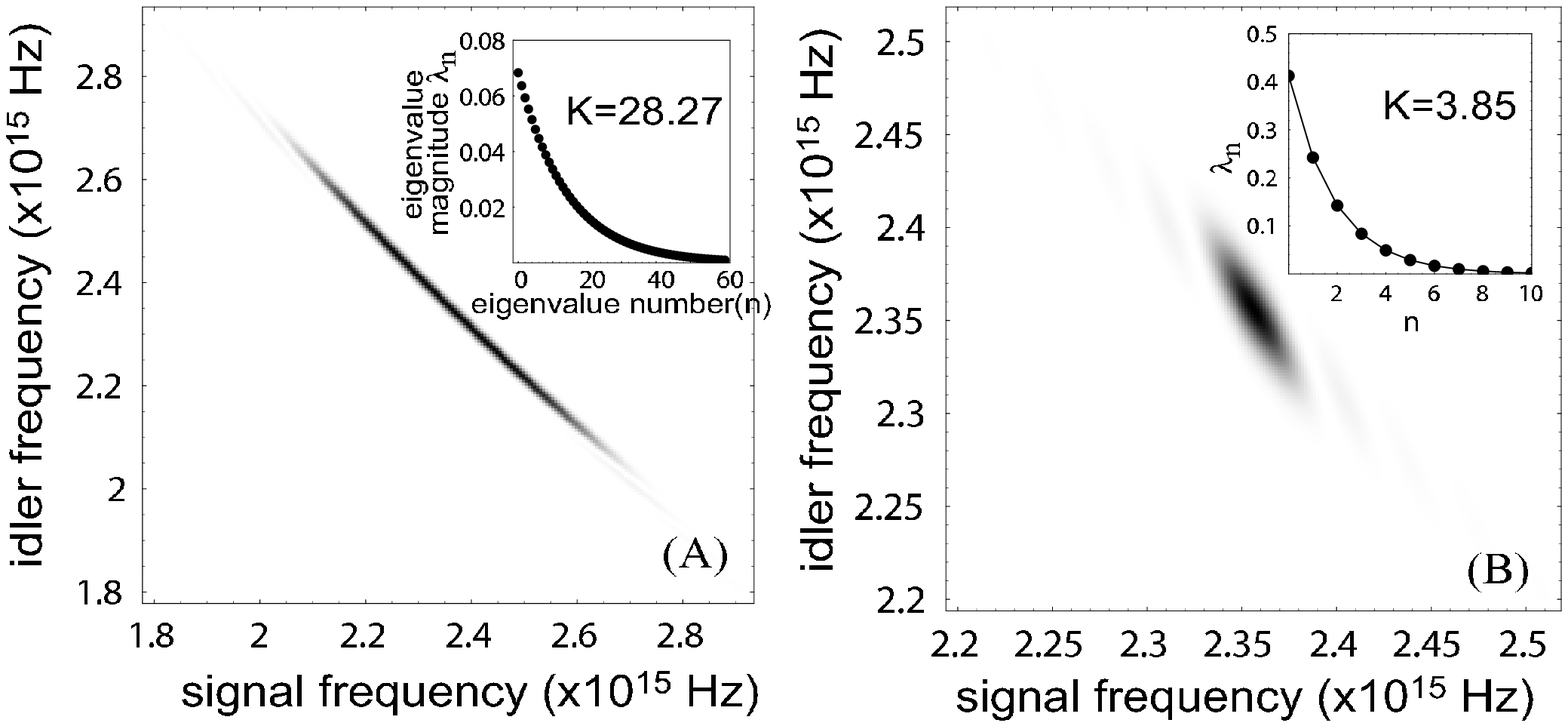, width=8.2cm}} 
\vspace*{13pt} \fcaption{\label{Fi:typ1and2} Joint spectral
intensity of ultrafast-pumped ($15$~nm bandwidth) PDC centered at
$800$~nm from a $1$~mm long BBO crystal.  (A) shows an example of
a two photon state involving non-collinear type I phase matching
and (B) shows a two photon state in the case of collinear type II
phase matching. Note that type I PDC has a higher degree of
spectral entanglement as quantified by the value of the
cooperativity parameter $K$. Such sources with high $K$ are not
suitable for a number of multiple-source experiments.}
\end{figure}

Let us turn our attention to actual examples of two-photon states
produced by PDC.   We will consider non-collinear type-I (eoo) and
collinear type-II (eoe) PDC processes in a beta-barium-borate
(BBO) crystal pumped by a train of ultrafast pulses.  Note that
for a Schmidt decomposition  of a PDC state to have physical
meaning the signal and idler photons must be distinguishable; in
the case of non-collinear type-I PDC this is ensured by distinct
spatial modes while in the case of type-II PDC this is ensured by
the orthogonal polarizations. For both of these cases, once we
select specific directions of propagation for the down-converted
photons, the joint two-photon probability amplitude becomes a
function of only the two frequencies $\omega_s$ and $\omega_i$ and
it can be written as follows \cite{WarrenPRA97}:
$S(\omega_s,\omega_i)=\alpha(\omega_s+\omega_i)
\phi(\omega_s,\omega_i)$ where $\alpha(\omega_s+\omega_i)$ is the
Fourier transform of the temporal pump amplitude and
$\phi(\omega_s,\omega_i)$ is the crystal phase matching function.
The pump envelope function may be modelled as a Gaussian
$\alpha(\omega_s+\omega_i)=\mbox{exp}[-
(\omega_s+\omega_i-2\omega_0)^2/\sigma_p^2]$ with a bandwidth
$\sigma_p$ and which is centered at $2\omega_0$.  The phase
matching function is given as
$\phi(\omega_s,\omega_i)=\mbox{sinc}(L \Delta k/2)$ where $L$ is
the crystal length and $\Delta
k=k_p(\omega_p)-k_s(\omega_s)-k_i(\omega_i)$ incorporates the
dispersion characteristics of the crystal. In
Fig.~\ref{Fi:typ1and2} we see that for typical experimental
parameters both type I and type II PDC, though particularly the
former, exhibit large spectral entanglement, as quantified by the
large value of $K$.

Is a high value of $K$ such as is obtained from a typical PDC
crystal desirable?  The answer to this question depends on what
experiment we  wish to perform.  In the following two sections we
will explore the specific requirements on two-photon states
necessary to guarantee high visibility interference in two cases:
i) multiple-crystal interferometry experiments and ii) experiments
relying on polarization entangled photon pairs.

\subsection{Criteria for the design of photon states to be used in
multiple-crystal interferometry.}\label{Sec:criteriamult}

A basic requirement necessary for the successful implementation of
many photon-based quantum computation protocols is the
availability of multiple-photon entangled states. In this section
we explore the experimental challenges likely to be faced when
synthesizing such a multiple-photon state from a number of
two-photon sources pumped synchronously with an ultrashort pulsed
pump. Employing such an approach, distinguishing information
implicit in timing can be limited to the pulsed pump temporal
duration. Unfortunately, distinguishing information may remain in
the other degrees of freedom. Filtering is a convenient way often
used in experiments to accurately define spatio-temporal modes
from PDC crystals and therefore eliminate distinguishing
information. Filtering, however, unfortunately leads to a
prohibitive reduction of the usable photon production rate.
Eliminating  spectral distinguishing information \textit{without
resorting to filtering}, in the context of multiple-crystal
interferometry, is the theme of this section. That filtering is
undesirable is manifest in that a multiple photon source based on
several PDC sources necessitates the simultaneous emission of
photon pairs from all sources used, which unfortunately occurs
with an exponentially decreasing probability as the number of
sources is increased (if the probability of obtaining a random
event from one source is $p$ the probability of $N$ simultaneous
events from $N$ sources is $p^N$). Though it is rather difficult
to view such an approach based on PDC as a practical route towards
fully scalable all-optical quantum computation, it can be
reasonably expected that this process can be applied in
small-scale quantum circuits used for example in quantum
communications.

Solid state sources of single photons have recently been reported
to exhibit enough indistinguishability for quantum interference
effects to be observed\cite{YamamotoNature}.  Such sources,
however, require a relatively complex apparatus including the need
for cryostatic cooling.  On the other hand, it is difficult to
direct emission to well-defined spatial modes, making collection
efficiencies of the produced photons quite limited. PDC-based
sources of indistinguishable photons, to be discussed in this
section, operate at room temperature and require little besides a
$\chi^{(2)}$ crystal, focused pump beam and apertures to define
the output spatial modes.

Under what conditions are photons emitted by different sources
indistinguishable?  The well-known Mandel dip experiment
\cite{Hong87} is a good platform on which to test the concept of
indistinguishability. For the classical Hong-Ou-Mandel
interferometer (HOMI) it turns out that the fourth-order
interference visibility depends only on the degree to which the
joint spectral amplitude is symmetric [symmetry defined for our
purposes as $S(\omega_s,\omega_i)=S(\omega_i,\omega_s)$ where
$S(\omega_s,\omega_i)$ is the joint spectral amplitude], assuming
that the minimum of the interference dip occurs for zero time
delay between the interfering photons. The physical reason for
this is that the two Feynman paths contributing to a coincidence
event ($R_xR_x$ path: $\omega_a$ in the idler arm reflected and
$\omega_b$ in the signal arm reflected. $T_xT_x$ path: $\omega_a$
in the signal arm transmitted and $\omega_b$ in the idler arm
transmitted, where $\omega_a$ and $\omega_b$ are the frequencies
of the photons registered by the respective detectors) become
distinguishable if the symmetry condition is not fulfilled. Thus,
for the Mandel dip experiment a large value of $K$ does not
constitute a limitation, as indeed has been shown experimentally
by the nearly $100\%$ interference visibility obtained for type-I
PDC in several experiments both in the CW and pulsed pumped
regimes.

\begin{figure} [htbp]
\vspace*{13pt}
\centerline{\psfig{file=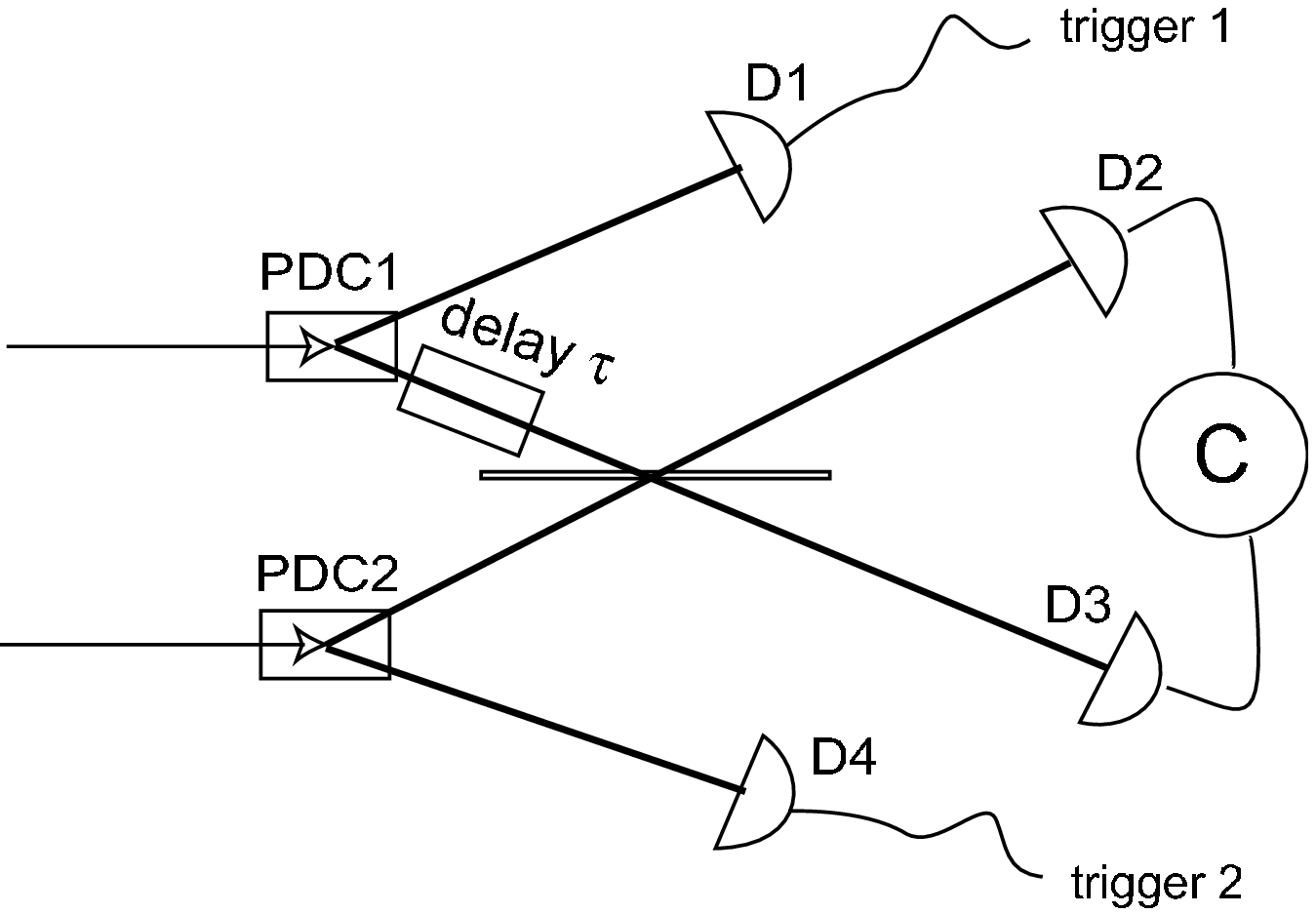, width=6.2cm}} 
\vspace*{13pt} \fcaption{\label{Fi:2xtalHOMI} Two PDC crystal HOMI
arrangement.  This is an example of an experiment where
interference visibility is reduced if the photon pairs are
spectrally correlated. Spectral correlation leads to
distinguishing information between the two photons which inhibits
interference.}
\end{figure}

Let us now discuss a two-PDC crystal interference experiment in an
event-ready HOMI-like arrangement. Consider two PDC crystals as
shown in Fig.~\ref{Fi:2xtalHOMI} and a HOMI arrangement
interfering the idler from the first crystal with the signal from
the second crystal, the remaining two channels being used as
triggers.  The experiment would consist of monitoring quadruple
coincidences on $D1$ through $D4$ while scanning the delay $\tau$.
If both PDC crystals are identical and described by a joint
spectral amplitude function $f(\omega_s,\omega_i)$, the four-fold
coincidence rate $R_c(\tau)$ may be shown to be proportional to
\cite{WarrenThesis}:
\begin{equation}\begin{split}
R_c(\tau)\propto
\int\limits_0^\infty\int\limits_0^\infty\int\limits_0^\infty\int\limits_0^\infty
d\omega_1 d\omega_2 d\omega_3 d\omega_4
&f(\omega_1,\omega_2)f(\omega_3,\omega_4)[
f^*(\omega_1,\omega_2)f^*(\omega_3,\omega_4)\\
&-e^{i(\omega_1-\omega_3)
\tau}f^*(\omega_1,\omega_4)f^*(\omega_3,\omega_2)]
\label{E:2xtalRc} \end{split}.
\end{equation}

We can see from Eq.~(\ref{E:2xtalRc}) that the visibility becomes
unity if at $\tau=0$, i.e. at zero delay between the two
interferometer paths, the coincidence rate vanishes.  A sufficient
condition for this to occur is that the integrand itself vanish at
$\tau=0$, which leads to the condition

\begin{equation}
f(\omega_1,\omega_2)f(\omega_3,\omega_4)=f(\omega_1,\omega_4)f(\omega_3,\omega_2)
\label{E:2xtalcond}
\end{equation}

for any $\omega_1$, $\omega_2$, $\omega_3$ and $\omega_4$.

Unlike in the one-crystal Mandel dip experiment, symmetry in the
function $f(\omega_s,\omega_i)$ does not help in fulfilling this
condition.   However, it may be seen that such a condition is met
if the function $f(\omega_s,\omega_i)$ is factorizable i.e. if
functions $p(\omega)$ and $q(\omega)$ exist such that
$f(\omega_s,\omega_i)=p(\omega_s)q(\omega_i)$.   For such a
factorizable state there is a single Schmidt mode pair i.e. $K=1$,
which means that there is no spectral entanglement.  The nature of
distinguishing information is therefore different in the case of
multiple PDC source experiments to that of single source
experiments. Whereas in the former distinguishing information
resides in the spectral entanglement, in the latter it resides in
the joint spectral amplitude asymmetry.

By what physical mechanisms can we obtain such a factorizable
state? Perhaps the simplest way is by making use of spectral
and/or spatial filtering. A recent experiment by Mlynek {\em et
al.}\ \cite{Mlynek} demonstrates the way in which distinguishing
information can be eliminated via filtering but at the cost of
extremely low resulting production rates.  To understand how this
works, consider a simplified type-I source where the joint
spectral amplitude is approximated by a Gaussian (i.e. we neglect
the secondary peaks of the sinc function and also neglect the
effects of dispersion) and is multiplied by a Gaussian spectral
filter function:
\begin{equation}
f(\nu_s,\nu_i)=A\mbox{ }
\mbox{exp}\left[-2(\nu_s^2+\nu_i^2)\left(\frac{1}{\sigma_F^2}+\frac{1}{\sigma^2}\right)-\frac{2
\nu_s \nu_i}{\sigma^2}\right] \label{E:ModelSource}
\end{equation}
where $A$ is a normalization constant, $\nu_j=\omega_j-\omega_0$
[with $j=s,i$], $\sigma_F$ is the width of the Gaussian spectral
filter and $\sigma$ is a resulting width of the product of the
pump envelope and phase matching functions. Substituting
Eq.~(\ref{E:ModelSource}) into Eq.~(\ref{E:2xtalRc}) and carrying
out the integrals we obtain the coincidence rate function:
\begin{equation}
R_c(\tau) \propto R_0\left[1-V\mbox{exp}\left(-\frac{\sigma^2
\sigma_F^2 \tau^2}{8(\sigma_F^2+\sigma^2)}\right)\right]
\label{E:Rc}
\end{equation}
which represents an interference dip with a temporal width
proportional to $\sqrt{\sigma^2+\sigma_F^2}/(\sigma \sigma_F)$ and
a visibility $V$ given by:

\begin{equation}
V=\sqrt{1-\frac{\sigma_F^4}{(\sigma_F^2+\sigma^2)^2}}
\label{E:visibility}
\end{equation}
The normalized count rate, $R_0$ (i.e. the coincidence rate in the
limit $\tau\rightarrow\infty$) is given by:
\begin{equation}
R_0=\frac{2 \sigma_F^2}{2\sigma_F^2+\sigma^2} \label{E:R0}
\end{equation}
which has a value between $0$ and $1$. For a fixed value of
$\sigma$, the filtering strength determines the value of $R_0$,
which may be seen to vanish for strong filtering
($\sigma_F\ll\sigma$) and gives unity in the limit of no filtering
($\sigma_F\gg\sigma$).

\begin{figure} [htbp]
\vspace*{13pt}
\centerline{\psfig{file=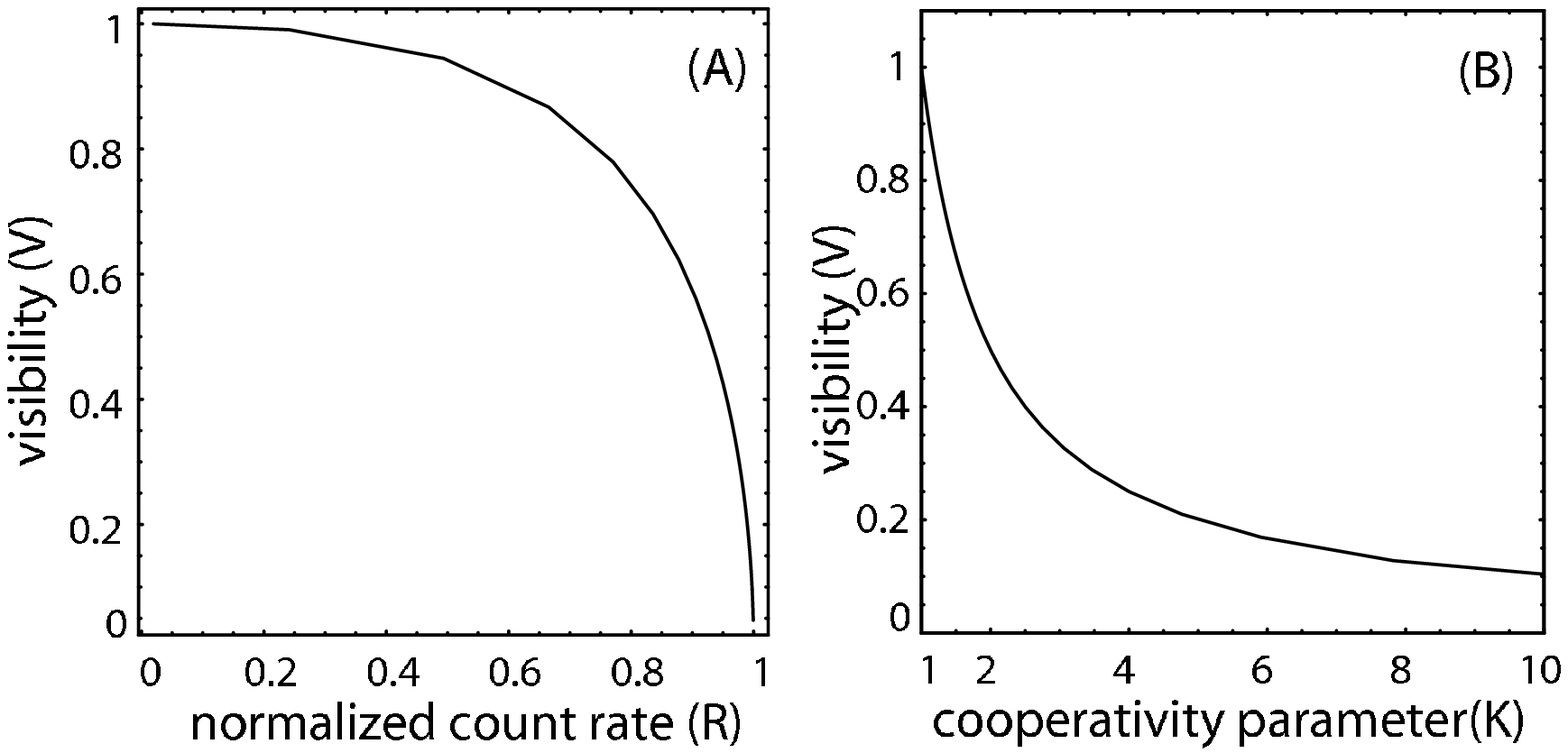, width=8.2cm}} 
\vspace*{13pt} \fcaption{\label{Fi:visKvisR} (A) depicts the
tradeoff between the interference visibility and the count rate
expected in a two-crystal HOMI setup for different filtering
strengths. Unit visibility is obtained only in the limit where the
filtering bandwidth approaches zero. (B) depicts the relationship
between the number of active Schmidt modes, the cooperativity
parameter, and the visibility. Interference is reduced if there is
more than one mode. In generating this plot
Eqs.~(\ref{E:Kanalytical}) and (\ref{E:mumodelsource}) from the
appendix were used.   For the plots in this figure, a type-I
simplified source with $\sigma=4 \times 10^{13} Hz$ was assumed
while the value of $\sigma_F$ was allowed to take a range of
values.}
\end{figure}

In the strong filtering regime (i.e. $\sigma_F\ll\sigma$),
Eq.~(\ref{E:visibility}) predicts that the visibility reaches
unity. In this case the joint spectral amplitude reduces to
$f(\nu_s,\nu_i)=\mbox{exp}[-2(\nu_s^2+\nu_i^2)/\sigma_F^2]$, which
is factorizable [and thus fulfils Eq.~(\ref{E:2xtalcond})] with
identical $p(\nu)$ and $q(\nu)$ functions given by
$p(\nu)=q(\nu)=\mbox{exp}(-{2 \nu^2}/{\sigma_F^2})$. Spectral
filtering can be straightforwardly carried out experimentally with
suitable interference filters readily available. Filtering,
however, has the unfortunate consequence that the count rate is
prohibitively reduced by rejecting most of the photon pairs (and
retaining only those without distinguishing information). This is
illustrated in Fig.~\ref{Fi:visKvisR}(A) which shows the
relationship between the visibility [see Eq.~(\ref{E:visibility})]
in the two-crystal HOMI and the expected count rate [see
Eq.~(\ref{E:R0})] with each point along the curve corresponding to
a different filtering strength. Fig.~\ref{Fi:visKvisR}(B) depicts
the relationship between the expected two-crystal visibility and
the cooperativity parameter [assuming that both PDC sources are
characterized by the simplified type-I spectral amplitude given in
Eq.~(\ref{E:ModelSource})]. Note that unit visibility is obtained
only in the limiting case of a single active Schmidt mode $K=1$,
that is in the case where each PDC source exhibits no spectral
correlation.

\subsection{Criteria for the design of polarization entangled
photon pairs}\label{Sec:criteriapolent}

The polarization of light fields can be readily controlled using,
for example, wave plates and polarizers.  This is in marked
contrast with the case of other photonic degrees of freedom, such
as frequency, for which it is rather more difficult to achieve a
comparable level of control.  It is therefore not surprising that
most recent experiments exploring issues of entanglement have
relied on polarization entanglement.  A number of reliable methods
for generating polarization entangled photon pairs have been
proposed and implemented \cite{Kwiat95,Kwiat99}.

As in any interference experiment, a pre-condition for high
interference visibility is that distinguishing information between
the interfering pathways is eliminated, even in those degrees of
freedom which are not of primary interest for the experiment in
question.  This applies to the generation of
polarization-entangled photon pairs, where even though
polarization is the degree of freedom of interest, it is crucial
to appropriately engineer the spectral properties of the photon
pairs to eliminate any spectral distinguishing
information\cite{RSpaper}.

\begin{figure} [htbp]
\vspace*{13pt}
\centerline{\psfig{file=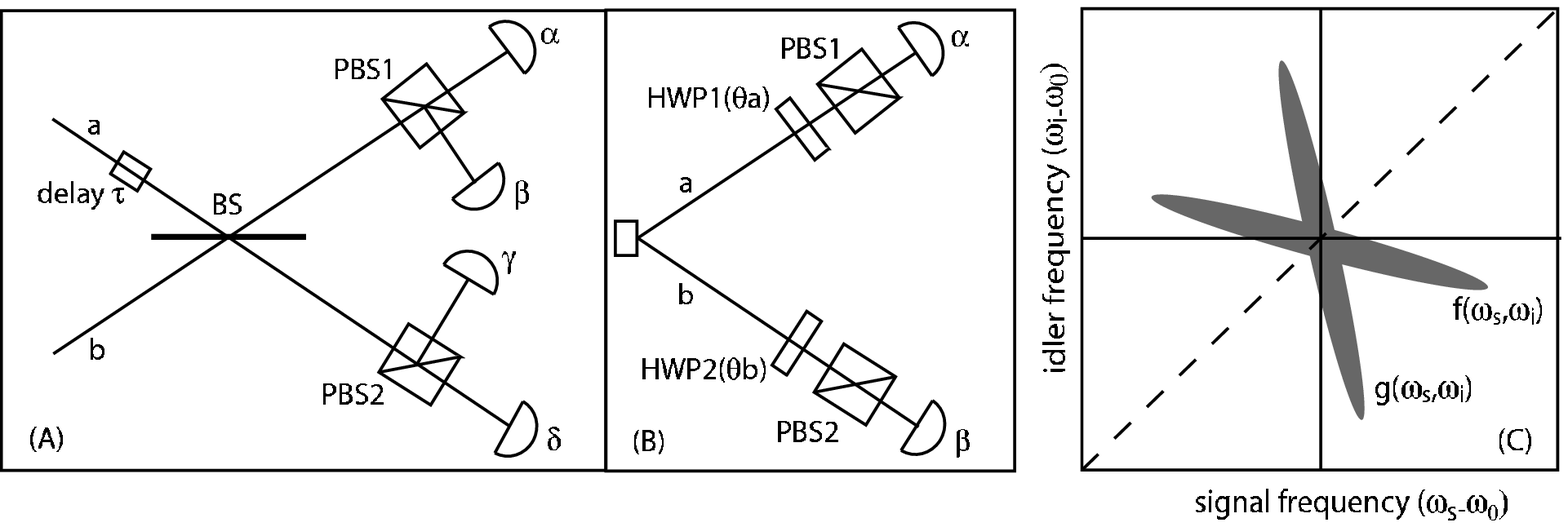, width=10.2cm}} 
\vspace*{13pt} \fcaption{\label{Fi:fandg} (A) Braunstein-Mann Bell
state analyzer. (B) Polarization correlation experiment.
(C)Graphical representation of the condition for optimal
performance.}
\end{figure}

We will illustrate the importance of this point by studying the
Braunstein-Mann Bell state analyzer \cite{BraunsteinMann}. The
apparatus is displayed in figure~\ref{Fi:fandg}(A) where BS is a
50:50 beam-splitter and PBS1 and PBS2 are polarizing
beam-splitters. This experiment serves as a building block for a
large class of experiments relying on Bell state measurement, such
as entanglement swapping and teleportation.  Employing such an
analyzer, the experimenter can make inferences about the state of
the incoming photon pair from the firing pattern of the four
detectors. Likewise, knowing the state of the incoming photon
pair, inferences can be made about what the firing pattern will
be. For instance, assuming that there is zero temporal delay
between the two photons ($\tau=0$), if $\alpha$ and $\gamma$ or
$\beta$ and $\delta$ fire we then know with certainty that the
incoming state is the singlet state
$|\psi^{(-)}\rangle_{ab}=2^{-\frac{1}{2}}(|HV\rangle_{ab}-|VH\rangle_{ab})$
[where $H$ and $V$ refer to horizontal and vertical polarization
and $a$ and $b$ refer to the two spatial modes]. Likewise, if the
incoming state is
$|\psi^{(+)}\rangle_{ab}=2^{-\frac{1}{2}}(|HV\rangle_{ab}+|VH\rangle_{ab})$,
we can infer with certainty that either $\alpha$ and $\beta$ or
$\delta$ and $\gamma$ will fire.  Considering the fact that the
Bell state analyzer relies on quantum interference, we would
expect that if the two-photon state used exhibits spectral
distinguishing information, the correlation between the incoming
two-photon state and the firing pattern will no longer be perfect.
We will compare the behavior of the Bell-state analyzer with that
of the apparatus shown in Fig~\ref{Fi:fandg}(B), which measures
the polarization correlations present in a polarization-entangled
two-photon state.

We begin our analysis by considering a maximally-entangled
polarization Bell state with the spectral component of the state
included in explicit form:

\begin{equation}
|\psi^\pm\rangle=\frac{1}{\sqrt{2}}\int\limits_0^\infty\int\limits_0^\infty
d\omega_1d\omega_2\left[f(\omega_1,\omega_2)a^\dag_H(\omega_1)b^\dag_V(\omega_2)\pm
g(\omega_1,\omega_2)a^\dag_V(\omega_1)b^\dag_H(\omega_2)\right]|\mbox{vac}\rangle
\label{E:psipm}
\end{equation}

where $a$ and $b$ refer to two distinct spatial modes, and where
the functions $f(\omega_1,\omega_2)$ and $g(\omega_1,\omega_2)$
are each normalized such that the integral over both arguments of
the modulus squared of the function is unity.  In order to study
the performance of the Bell state analyzer, we detect the whole
output from each port of the beam-splitter in
figure~\ref{Fi:fandg}(A) i.e. we remove the two polarizing
beamsplitters and place a detector in each output mode.   If the
Bell state analyzer operates ideally, as described above, the
singlet $|\psi^{(-)}\rangle$ state yields a unit coincidence rate
while the $|\psi^{(+)}\rangle$ yields a zero coincidence rate.  On
the other hand, calculating the coincidence rate assuming a
general incoming state given as in eq.~(\ref{E:psipm}), we obtain
the result:

\begin{equation}
R_c^\pm(\tau)=\frac{1}{4}\int\limits_0^\infty\int\limits_0^\infty
d\omega_1
d\omega_2\left|f(\omega_1,\omega_2)\mp\mbox{e}^{i(\omega_1-\omega_2)\tau}
g(\omega_2,\omega_1)\right|^2 \label{E:RcBell}
\end{equation}

where the $+/-$ on the left hand side refers to a
$\psi^{(+)}/\psi^{(-)}$ incoming state. Because the integrand in
eq.~(\ref{E:RcBell}) is non-negative, the necessary and sufficient
condition to obtain $R_c^+=0$ and $R_c^-=1$ at $\tau=0$ (i.e. the
desired behavior) is that:

\begin{equation}
g(\omega_1,\omega_2)=f(\omega_2,\omega_1) \label{E:fgcond}
\end{equation}

\begin{figure} [htbp]
\vspace*{13pt}
\centerline{\psfig{file=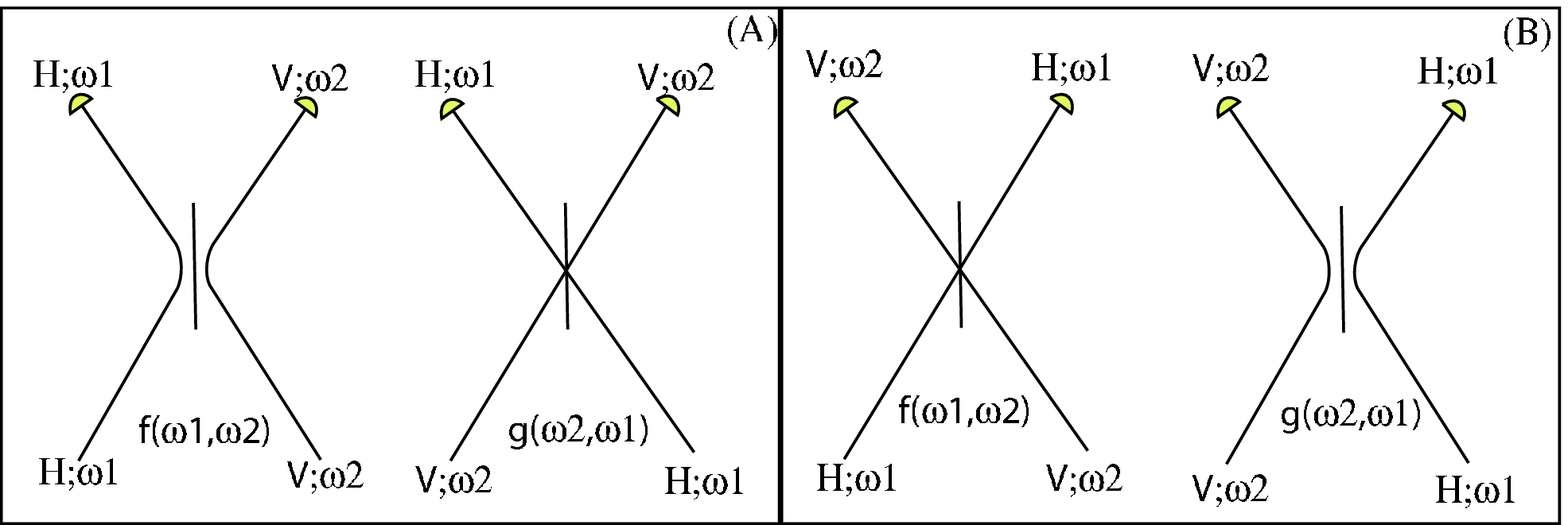, width=10.2cm}} 
\vspace*{13pt} \fcaption{\label{Fi:feynman} The two Feynman
alternatives shown in each box are indistinguishable from each
other.}
\end{figure}

Fig.~\ref{Fi:fandg}(C) depicts the condition in
eq.~(\ref{E:fgcond}) graphically: the two spectral amplitude
functions must be the specular image of each other across the line
$\omega_1=\omega_2$. In order to gain a physical understanding of
this condition we consider the Feynman alternatives leading to
obtaining a coincidence event. Suppose that the photon pair is
described by the first component of the Bell state, i.e. the
amplitude containing $f(\omega_1,\omega_2)$ and that the two
photons are reflected at the beam splitter. Such an event [left
diagram in Fig.~\ref{Fi:feynman}(A)] is indistinguishable (meaning
having an identical detection pattern) from the event where the
photon pair is described by the second amplitude with reversed
frequency arguments i.e. $g(\omega_2,\omega_1)$, and where the two
photons are transmitted at the beam splitter [see right diagram in
Fig~\ref{Fi:feynman}(A)].  For this system, there is a second set
of two pathways as shown in Fig.~\ref{Fi:feynman}(B) which may
lead to a coincidence. For unit interference visibility to occur,
the two pathways in each set must be indistinguishable from each
other. Fig.~\ref{Fi:feynman} shows schematically that for both
sets of pathways, indistinguishability is guaranteed (for
$\tau=0$) if the condition in Eq.~(\ref{E:fgcond}) is fulfilled.

Consider now the apparatus in Fig.~\ref{Fi:fandg}(B), which is the
standard method used has been used to test whether a photon pair
is polarization-entangled. Assuming that the incoming state is
given as in eq.~(\ref{E:psipm}),  the coincidence rate as a
function of the polarization rotation angles in each arm,
$\theta_a$ and $\theta_b$, may be shown to be given by:

\begin{equation}
R_c^\pm(\theta_a,\theta_b)=\int\limits_0^\infty\int\limits_0^\infty
d\omega_1 d\omega_2 \left|\mbox{cos}\theta_a \mbox{sin}\theta_b
f(\omega_1,\omega_2)\pm \mbox{sin}\theta_a \mbox{cos} \theta_b
g(\omega_1,\omega_2)\right|^2 \label{E:polcorr}
\end{equation}

The resulting necessary and sufficient condition to obtain unit
visibility fringes [in which case the coincidence rate reduces to
$R_c^\pm(\omega_a,\omega_b)=\mbox{sin}^2(\theta_a\pm\theta_b)$]
is:
\begin{equation}
f(\omega_1,\omega_2)=g(\omega_1,\omega_2) \label{E:polcorrcond}
\end{equation}

Note that this condition, which says that the two amplitudes
should be identical, is different from that derived for the Bell
state analyzer experiment [see eq.~(\ref{E:fgcond})]. The fact
that these conditions differ means that depending on the kind of
experiment to be performed the polarization-entangled state to be
used should entail differently engineered spectral properties.

Is it possible to fulfil both conditions simultaneously? Together,
the two conditions reduce to the requirement that the quantum
amplitudes be identical and symmetric i.e.
$f(\omega_1,\omega_2)=f(\omega_2,\omega_1)=g(\omega_1,\omega_2)$.
Generating a polarization-entangled state using two type-I
cascaded crystals\cite{Kwiat99}, the two amplitudes
$f(\omega_1,\omega_2)$ and $g(\omega_1,\omega_2)$ are each
symmetric and can be made identical, even in the case where a
pulsed pump is used, if the pump is pre-compensated as described
by Nambu {\it et al.}\cite{Nambu} .  Therefore, this method for
generating polarization-entangled states has the important
property that it can be made to fulfil both conditions
[Eq.~(\ref{E:fgcond}) and Eq.~(\ref{E:polcorrcond})], so that the
same source may be used successfully (without the need for
spectral filtering) for the polarization-correlation experiment
[Fig.~\ref{Fi:fandg}(A)] and for the Bell state analyzer
experiment [Fig.~\ref{Fi:fandg}(B)].

Polarization-entangled photon pairs may also be synthesized using
the modes defined by the intersections of the ``e''-ray and
``o''-ray cones in type II PDC\cite{Kwiat95}. It may be shown that
this scheme produces a polarization-entangled state such that the
condition in eq.~(\ref{E:fgcond}) is fulfilled. This means that
such a state is well-suited for Bell state-measurement experiments
[e.g. the Bell state analyzer in Fig.~\ref{Fi:fandg}(A)] while it
is not well suited for the polarization correlation experiment
[Fig.~\ref{Fi:fandg}(B)]. In a recent experiment making use of
this type II source \cite{KimGricePRA, KimGriceJMO} , one of the
two spatial modes is subjected to a $90^\circ$ polarization
rotation prior to be being combined at a beamsplitter with the
other mode. The authors show that the two output modes from the
beamsplitter are in a polarization-entangled state which yields
near unit visibility, without resorting to spectral filtering, in
a polarization correlation experiment such as the one shown in
Fig.~\ref{Fi:fandg}(B).  We can understand the results of Kim
\textit{et al.} in terms of the conditions in Eq.~(\ref{E:fgcond})
and Eq.~(\ref{E:polcorrcond}). The half-wave plate and
beamsplitter accomplish turning a state obeying
$f(\omega_1,\omega_2)=g(\omega_2,\omega_1)$ into a state obeying
$f(\omega_1,\omega_2)=g(\omega_1,\omega_2)$.  The resulting state,
however, is no longer well suited for Bell-measurement
experiments.

For a polarization-entangled photon pair based on type-II PDC to
be simultaneously optimized for both kinds of experiments (Bell
state analysis and polarization correlation), the signal and idler
photons need to fulfil the so-called group velocity matching
condition \cite{RubinKeller, WarrenAlfred} which for a given
$\chi^{(2)}$ material is met at a specific (usually long, $>1 \mu
\mbox{m}$) wavelength.   For BBO, for example, such a condition is
met at a central PDC wavelength of $1.51 \mu \mbox{m}$ at which
single-photon detection is not technologically well-developed.  It
would therefore appear that it is possible to optimize an
experimentally-feasible polarization-entangled photon pair
generated via type-II PDC (which does not fulfil the group
velocity matching condition) for one of the two types of
experiments discussed above, but not for both simultaneously.
There are in fact two ways in which the amplitudes may be made
identical and symmetric, and therefore be made to fulfil both
conditions. The first is by pumping the type-II source with a CW
pump: in the limit of zero pump bandwidth, the pump envelope
function which is always symmetric dominates over the phase
matching function to determine the overall spectral amplitude. It
is also possible to symmetrize the amplitudes by placing a
(symmetric) interference filter with a narrow enough bandwidth and
an appropriately chosen central bandpass frequency at each of the
two spatial modes.  The latter, of course, at the cost of a sharp
reduction in the production rate of photon pairs. We therefore
conclude that an ultrafast-pumped spectrally unfiltered and
non-group-velocity-matched type-II polarization-entangled source
cannot be simultaneously optimized for both kinds of experiments,
in contrast with a two type-I cascaded crystal source.

As discussed above, a sufficient condition for a
polarization-entangled state to be simultaneously optimized for
Bell state measurement and for polarization correlation
experiments (see Fig.~\ref{Fi:fandg}) is that the joint spectral
amplitude be symmetric i.e.
$f(\omega_1,\omega_2)=f(\omega_2,\omega_1)$. This requirement is
fulfilled in the specific case where the joint spectral amplitude
exhibits no spectral correlations (see also
Sec.~\ref{Sec:nocorr}). Here we investigate the possible
advantages of engineering polarization entangled two-photon states
so that they are also spectrally uncorrelated.

Why would such a state be useful?  In order to answer this
question we analyze the $|\phi^\pm\rangle$ state such as would be
produced by a cascaded type-I polarization-entangled PDC
source\cite{Kwiat99,Nambu}:

\begin{equation}
|\phi^\pm\rangle=\frac{1}{\sqrt{2}}\int\limits_0^\infty\int\limits_0^\infty
d\omega_1d\omega_2 f(\omega_1,\omega_2)
\left[a^\dag_H(\omega_1)b^\dag_H(\omega_2)\pm
a^\dag_V(\omega_1)b^\dag_V(\omega_2)\right]|\mbox{vac}\rangle
\label{E:phipm}
\end{equation}

where the joint spectral amplitude is assumed to be symmetric:
$f(\omega_1,\omega_2)=f(\omega_2,\omega_1)$. Suppose now that the
spectral amplitude is uncorrelated (i.e. has a cooperativity
parameter value $K=1$) such that functions $p(\omega)$ and
$q(\omega)$ exist yielding
$f(\omega_1,\omega_2)=p(\omega_1)q(\omega_2)$. The two-photon
state [Eq.~(\ref{E:phipm})] becomes:

\begin{equation}
|\phi^\pm\rangle=\frac{1}{\sqrt{2}}(c^\dag_H d^\dag_H \pm
c^\dag_Vd^\dag_V) |\mbox{vac}\rangle \label{E:simpphipm}
\end{equation}

where:
\begin{align}
c^\dag_{H,V}&=\int\limits_0^\infty d\omega p(\omega)a^\dag_{H,V}(\omega)\label{E:cHV}\\
d^\dag_{H,V}&=\int\limits_0^\infty d\omega
q(\omega)b^\dag_{H,V}(\omega) \label{E:dHV}
\end{align}

We see from Eq.~(\ref{E:simpphipm}) that a spectrally uncorrelated
polarization-entangled two-photon state can be written in terms of
``effective'' creation operators, in which the spectral degree of
freedom no longer plays a role once integrated out [see
Eq.~(\ref{E:cHV}) and Eq.~(\ref{E:dHV})]. The fact that the
spectral degree of freedom is no longer present means that
spectral indistinguishability is guaranteed in any interference
experiment, making this an ideal source of polarization entangled
photons. Operators $a^\dag_{H,V}$ corresponding to spatial mode
$a$ are translated into effective operators $c^\dag_{H,V}$ [see
Eq.~(\ref{E:cHV})] while operators $b^\dag_{H,V}$ corresponding to
spatial modes $b$ are translated into effective operators
$d^\dag_{H,V}$ [see Eq.~(\ref{E:dHV})]. We have thus shown that if
polarization-entangled photon pairs are appropriately engineered,
the spectral degree freedom can be altogether eliminated.

The use of the spectrally engineered polarization-entangled photon
pairs described in this section together with stimulated PDC as
described by De Martini and co-workers in Ref.\cite{DeMartini} and
Lamas-Linares and co-workers in Ref.\cite{LamasLinares}, opens up
the possibility of generating photon pairs exhibiting more complex
kinds of entanglement while maintaining spectral
indistinguishability.  The availability of such states may be of
crucial importance for realizing quantum computation schemes.

\section{Engineering photon pairs for quantum
information.}\label{Sec:eng}

\subsection{Engineering states with no spectral
correlations}\label{Sec:nocorr}

Our approach is to investigate ways to engineer the state
\textit{at the source} in order to obtain a factorizable state
thus eliminating the need for filtering.  One such approach is
described in Ref.~\cite{WarrenAlfred} where it is shown that for
degenerate collinear type-II phase matching, and for an
appropriate choice of material, central PDC wavelength, crystal
length and pump bandwidth, it is indeed possible to obtain such a
spectrally uncorrelated state.  This scheme makes use of the fact
that in some spectral regions and for some materials the phase
matching function contours have a positive slope that can
counteract the negative slope of the pump envelope function.
Unfortunately, for most common $\chi^{(2)}$ materials, this
positive slope occurs at longer wavelengths ($>1~\mu$m), where
single-photon detectors are not technologically well developed.
For example in the case of BBO an uncorrelated state is possible
for central PDC wavelengths between $1.15~\mu$m and $1.92~\mu$m.

\begin{figure} [htbp]
\vspace*{13pt}
\centerline{\psfig{file=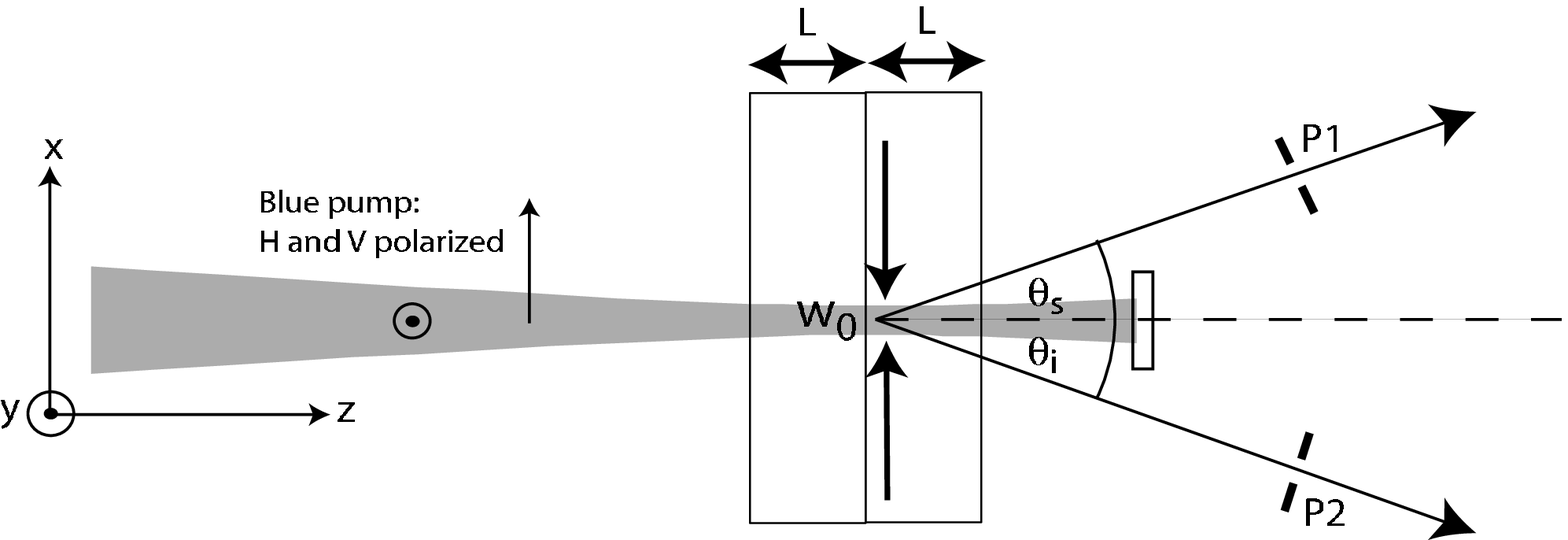, width=8.2cm}} 
\vspace*{13pt} \fcaption{\label{Fi:focusing} Uncorrelated photon
pair generation through non-collinear type-I PDC in a bulk type-I
crystal making use of a Gaussian beam pump. Pinholes $P1$ and $P2$
are required in order to accurately define two spatial modes.}
\end{figure}

Here we present a novel method for generating spectrally
uncorrelated pairs making use of non-collinear, degenerate type-I
PDC in bulk crystals and which exploits the transverse momentum of
the photons. This approach requires the ability to specify
accurately the crystal length and the beam diameter at the beam
waist (i.e.\ the focusing strength) and requires the spatial modes
exiting the crystal to be accurately defined for example with
pinholes or fibers.  An important additional requirement is an
ultrafast pulsed pump.  Fig.~\ref{Fi:focusing} outlines the
experimental setup. Unlike the approach in
Ref.~\cite{WarrenAlfred}, this method can be made to work at any
PDC central wavelength where phase matching is possible, in
particular at those where silicon based single-photon counting
modules work efficiently.

We begin by extending the phase matching function to the case
where a Gaussian beam, rather than a plane wave, is used as pump
field. In this case the electric field amplitude is given by:
\begin{equation}
\alpha(x,y,z;k)=\frac{1}{1+\frac{2iz}{k
w_0^2}}\mbox{exp}\left(-\frac{x^2+y^2}{w_0^2(1+\frac{2iz}{k
w_0^2})} \right) e^{ikz} \label{E:gaussianpump}
\end{equation}
where $w_0$ is the beam diameter at the beam waist. Following the
procedure sketched out in Ref.~\cite{WarrenPRA97}, we obtain for
the phase matching function:
\begin{equation}
\phi(\Delta k_z,\Delta k^\bot)\propto
\mbox{exp}\left(-\frac{(\Delta k^\bot)^2 w_0^2}{4}
\right)\mbox{sinc}\left[\left(\frac{(\Delta
k^\bot)^2}{4k}-\frac{\Delta k_z}{2}\right)L\right]
\label{E:fullPM}
\end{equation}
where $\Delta k^\bot$ refers to the transverse momentum along the
$x$-$z$ plane (see Fig.~\ref{Fi:focusing}) so that it is a scalar
quantity.  Making the approximation $\mbox{sinc}(x)\approx
\mbox{exp}(-\gamma x^2)$ (with $\gamma=.193...$\fnm{a}\fnt{a}{The
value $\gamma=.193...$ is obtained by letting the Gaussian
function used have the same FWHM width as the sinc function being
approximated}) and the further approximation $w_0/L\gg
\sqrt{\gamma}(\Delta k^\bot/k)^2$ (or equivalently $w_0/L
\gg\sqrt{\gamma} \mbox{sin}^2 \theta$ where $\theta$ is the
propagation angle of the downconverted photons) which sets an
upper limit on the focusing strength, the transverse momentum and
longitudinal momentum contributions to the phase matching function
can be factorized: $\phi(\Delta k_z,\Delta
k_\bot)\propto\phi_z(\Delta k_z)\phi^\bot(\Delta k^\bot)$ where:
\begin{align}
\phi_z(\Delta k_z)&=\mbox{exp}\left(-\frac{\gamma
\Delta k_z^2 L^2}{4} \right)  \label{E:PMapproxz}\\
\phi^\bot(\Delta k^\bot)&=\mbox{exp}\left(-\frac{(\Delta k^\bot)^2
w_0^2}{4} \right) \label{E:PMapproxt}
\end{align}

The joint spectral amplitude can now be written as
$S(\omega_s,\omega_i)=\alpha(\omega_s+\omega_i)\phi_z(\Delta
k_z)\phi^\bot(\Delta k^\bot)$, where $\alpha(\omega_s+\omega_i)$
is the pump envelope function.

\begin{figure} [htbp]
\vspace*{13pt}
\centerline{\psfig{file=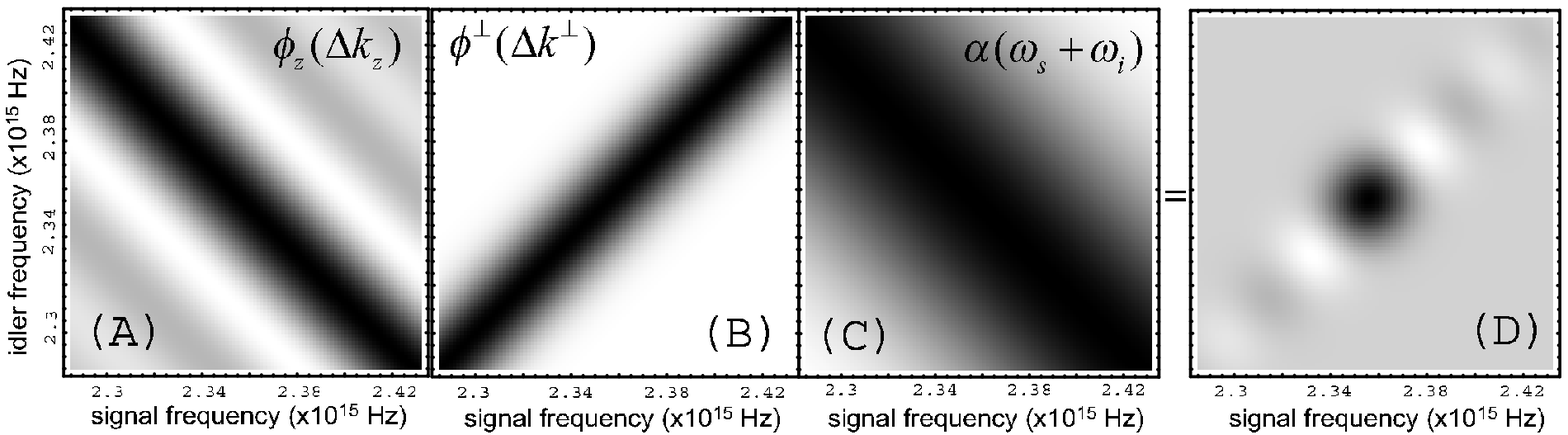, width=12.2cm}} 
\vspace*{13pt} \fcaption{\label{Fi:ZC} An uncorrelated two-photon
state can be synthesized by exploiting transverse momentum in the
crystal. Shown here is the case of a type-I BBO crystal pumped at
$400nm$ with output modes at $ \pm 3^\circ$.  (A) shows the
longitudinal phase matching function with negative unity contour
slope, (B) shows the transverse phase matching function with unit
contour slope, (C) shows the pump envelope function with a FWHM
bandwidth of $10$~nm, (D) shows the product of the three previous
functions, or the joint spectral amplitude, which is nearly
spectrally uncorrelated. We can see that a narrow bandwidth for
the pump envelope function would destroy the spectral
uncorrelation.}
\end{figure}

Let us restrict our attention to the case where the photons are
emitted along the $x$-$z$ plane (see Fig.~\ref{Fi:focusing}), so
that the direction of propagation of a given photon can be
described with the polar angle (defined within the crystal)
$\theta$ (and having zero azimuthal angle $\phi=0$). We assume
that the pump propagates such that $\theta_p=0$ in which case
conservation of transverse momentum dictates that
$\theta_s=-\theta_i=\theta$. Performing a Taylor expansion and
neglecting second and higher order terms, the longitudinal and
transverse phase mismatch are then given by:

\begin{align}
\Delta k_z&= \Delta k_z^{(0)}
+(k_p'-k'\mbox{cos}\theta)(\nu_s+\nu_i)
\label{E:deltakz}\\
\Delta k^\bot&=-k' \mbox{sin}\theta (\nu_s-\nu_i)\label{E:deltakt}
\end{align}

where $\nu_j=\omega_j-\omega_0$ with $j=s,i$.  All wavevector
amplitudes and their frequency derivatives are evaluated at
$\omega_0$ (or $2 \omega_0$ in the case of $k_p$ and $k_p'$).
$\Delta k_z^{(0)}=k_p-2 k \mbox{cos}\theta$ represents a constant
term which must vanish to ensure phase matching.  We ignore second
order and higher terms in Eqs.~(\ref{E:deltakz}) and
(\ref{E:deltakt}) since the resulting overall frequency range is
quite limited [see also Fig.~\ref{Fi:ZC}].  The central idea of
this approach is to exploit the fact that, as expressed in
Eqs.~(\ref{E:deltakz}) and (\ref{E:deltakt}), whereas the
longitudinal phase mismatch depends on the frequency sum
$\nu_s+\nu_i$, the transverse phase mismatch depends on the
frequency difference $\nu_s-\nu_i$.   This means that while the
contours of the longitudinal phase matching function
$\phi_z(\Delta k_z)$ have negative unit slope those of the
transverse phase matching function $\phi^\bot(\Delta k^\bot)$ have
positive unit slope.   Therefore, through an appropriate choice of
the widths of the two functions (proportional to $L^{-1}$ and
$w_0^{-1}$ respectively)  the overall phase matching function can
be made to be factorizable.  The physical reason behind the
positive slope of the transverse phase matching function is that
transverse momentum conservation leads to the signal and idler
photons propagating on opposite sides of the pump i.e.
$\theta_s=-\theta_i$.   Because this is a fundamental requirement
of non-collinear degenerate PDC, this scheme (unlike that in
ref.~\cite{WarrenAlfred}) works at any central PDC wavelength
where phase matching occurs.

When substituting Eq.~(\ref{E:deltakz}) into
Eq.~(\ref{E:PMapproxz}) and Eq.~(\ref{E:deltakt}) into
Eq.~(\ref{E:PMapproxt}), and multiplying the two resulting
functions, there is a cross-term proportional to $\nu_s \nu_i$ in
the exponential argument which is responsible for the
non-factorizability of the phase matching function.   Our approach
is to let this term vanish, thus imposing a constraint on the
crystal length $L$ and the beam diameter at the beam waist $w_0$.
This constraint tells the experimenter, for a given crystal
length, the required focusing strength needed in order to ensure a
factorizable phase matching function:
\begin{equation}
\frac{w_0}{L}=\frac{\sqrt{\gamma}(k_p'-k' \mbox{cos} \theta)}{k'
\mbox{sin} \theta} \label{E:Lw0}
\end{equation}
Note that there is a threshold bandwidth of the pump envelope
which must be exceeded for the spectral uncorrelation of the state
not to be destroyed by the strict spectral anti-corelation imposed
by a narrow (or CW) pump:
\begin{equation}
\sigma_p>\frac{\sqrt{2}}{\gamma L(k_p'-k'\mbox{cos} \theta)}
\label{E:pumpreq}
\end{equation}
where $\sigma_p$ refers to the pump bandwidth defined in Sec.2.
Let us look at a specific example involving a $1 mm$ long type-I
BBO crystal with a cut angle of $\theta_{PM}=30.32^\circ$. When
pumping such a crystal with an ultrafast pulsed pump centered at
$\lambda_p=0.4~\mu$m, degenerate non-collinear PDC is phase
matched at propagation angles (within the crystal) of
$\theta_s=3^\circ$ and $\theta_i=-3^\circ$.  The condition in
Eq.~(\ref{E:Lw0}) gives a beam diameter of $w_0=287~\mu$m.  Note
that with $w_0/L=0.287$ and $\theta=3^\circ$ we are in a regime
where the approximation $w_0/L \gg \sqrt{\gamma} \mbox{sin}^2
\theta$ is valid. Fig.~\ref{Fi:ZC} shows graphically the interplay
of the longitudinal and transverse phase matching functions which
combine to yield an uncorrelated state. Note that the
downconverted photons are in this example emitted with a central
wavelength of $\lambda=0.8~\mu$m so that they can be conveniently
detected with silicon based detectors.

\subsection{Engineering polarization entangled photon pairs without
spectral correlations}\label{Sec:polentnocorr}

\begin{figure} [htbp]
\vspace*{13pt}
\centerline{\psfig{file=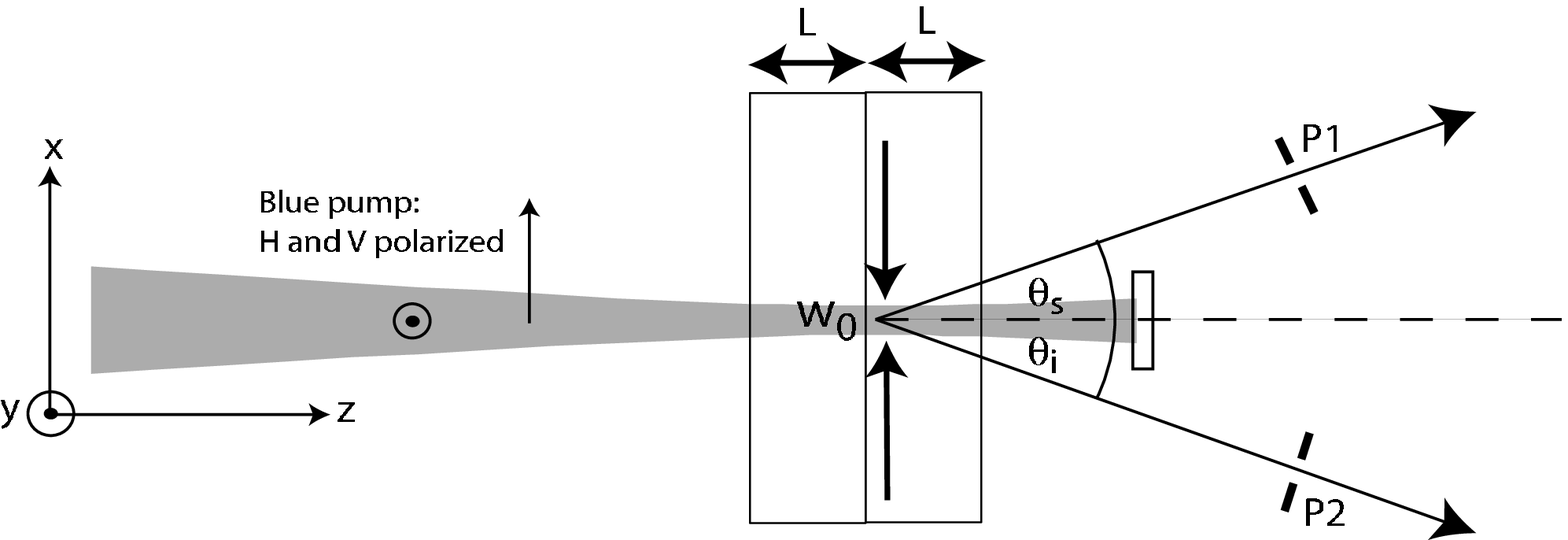, width=10cm}} 
\vspace*{13pt} \fcaption{Apparatus designed to generate a
polarization-entangled frequency-uncorrelated two-photon
pair.\label{Fi:ZCent} }
\end{figure}

As discussed at the end of Sec.~\ref{Sec:criteriapolent}, a
spectrally-uncorrelated polarization-entangled two-photon state
has an appealing structure in that the spectral degree of freedom
plays no role.  Here we propose a straightforward adaptation of
the technique described in the previous section for generating
such a spectrally-uncorrelated polarization entangled state.  This
method is based on the ultrafast-pumped cascaded type-I source
demonstrated by Nambu \textit{et al.}\cite{Nambu} and on the
technique designed to eliminate spectral correlations presented in
Sec.~\ref{Sec:nocorr}.  Two type-I crystals are placed
sequentially, with a relative optic axis rotation of $90^\circ$ as
shown in Fig.~\ref{Fi:ZCent}. A focused Gaussian pump beam is used
such that the crystal length (assumed to be the same for both
crystals) and beam waist diameter fulfil Eq.~(\ref{E:Lw0}). The
two blue pumps, polarized horizontally and vertically
respectively, have a relative temporal delay in order to
pre-compensate for group velocity mismatch as described by Nambu
and co-workers\cite{Nambu}.  The photon-pairs generated can be
written in terms of effective creation operators [see
Sec.~\ref{Sec:criteriapolent}] as in Eq.~(\ref{E:simpphipm}). The
state thus produced has the important property that spectral
indistinguishability is guaranteed in any interference experiment.

\subsection{Engineering photon pairs with coincident
frequencies}\label{Sec:poscorr}

There have been recent proposals for improved metrology by
exploiting quantum entanglement.  Giovannetti \textit{et
al.}\cite{GioNature} have proposed a quantum positioning scheme in
which the temporal uncertainty in the time of arrival of $N$
photons, and therefore the uncertainty in the inferred position of
the emitter with respect to the detectors, diminishes with the
number of photons $N$ by a factor $\sqrt{N}$ faster than is
possible with an analogous classical entanglement-free system.
This positioning scheme necessitates an $N$-photon ``frequency
correlated'' state where each of the emitted photons has an
identical frequency. In the case of $N=2$ such a state can be
written as:

\begin{equation}
|\Psi\rangle=\int\limits_0^\infty d\omega
f(\omega)|\omega\rangle_1|\omega\rangle_2
\label{E:poscorr}\end{equation}

where $f(\omega)$ gives the relative frequency weighting. Such
states, furthermore, exhibit an interesting dispersion
cancellation effect to all orders as described by Erdmann and
co-workers in Ref.\cite{ReinhardPRA} which opens up the exciting
possibility of transmitting qubits over arbitrarily long
dispersive fibers without decoherence.  A technique has been
proposed to generate such states as described by Giovanneti
\textit{et al.}\cite{GioPosCorr}.  This method relies on
satisfying the group velocity matching condition\cite{RubinKeller}
which unfortunately for most common $\chi^{(2)}$ materials occurs
at longer ($>1 \mu m$) wavelengths\cite{WarrenAlfred}.

\begin{figure} [htbp]
\vspace*{13pt}
\centerline{\psfig{file=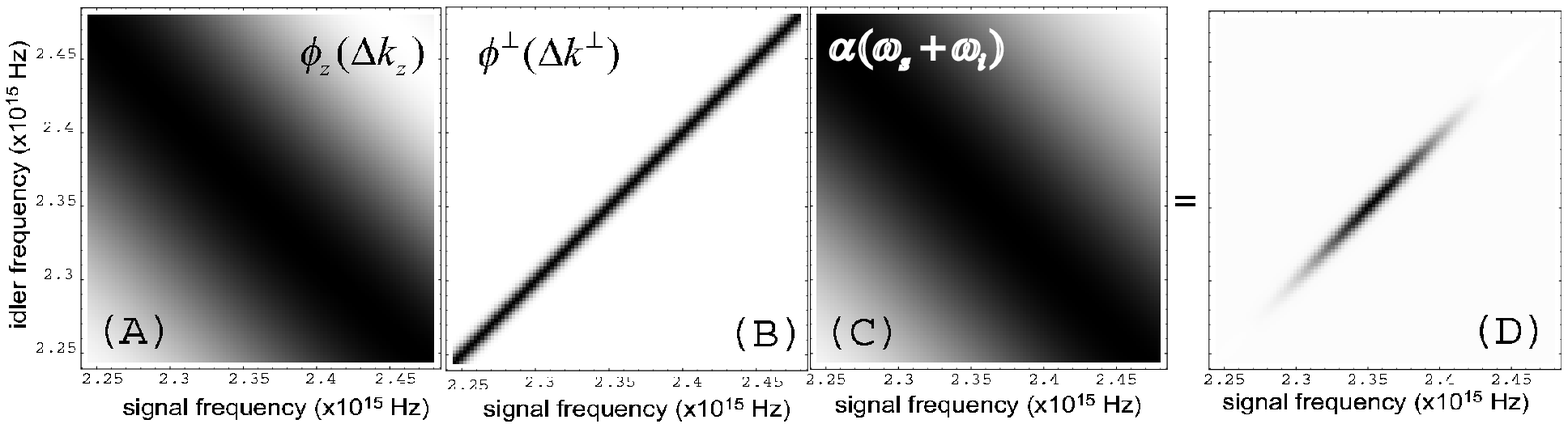, width=12.2cm}} 
\vspace*{13pt} \fcaption{\label{Fi:poscorr} A "frequency
correlated" two-photon state can be synthesized by exploiting
transverse momentum in the crystal. Shown here is the case of a
$200 \mu \mbox{m}$ long type-I BBO crystal pumped at
$400\mbox{nm}$ by a focused beam with spot size diameter
$w_0=1\mbox{mm}$ with output modes at $ \pm 3^\circ$. (A) shows
the longitudinal phase matching function with negative unity
contour slope, (B) shows the transverse phase matching function
with unit contour slope, (C) shows the pump envelope function with
a FWHM bandwidth of $15$~\mbox{nm}, (D) shows the product of the
three previous functions, or the joint spectral amplitude,
exhibiting frequency correlation.  Note that a strictly frequency
correlated state would have a vanishingly small width in the
$\nu_s-\nu_i$ direction.  The state shown in (D) represents a good
approximation to such an ideal state.}
\end{figure}

Here we propose a technique, based closely on the ideas described
in the previous two sections, to generate frequency correlated
states at wavelengths where Silicon-based single counting modules
are well-developed. Consider the technique proposed in
Sec.~\ref{Sec:nocorr} and shown schematically in Fig.~\ref{Fi:ZC}.
The basic idea behind this scheme is to let the longitudinal
phasematching function [see Eq.~(\ref{E:PMapproxz})] have a
relatively wide bandwidth while letting the transverse
phasematching function have a relatively narrow bandwidth [see
Eq.~(\ref{E:PMapproxt})]. A narrow-band \textit{longitudinal}
phasematching function can be obtained by using a short PDC
crystal while a wide-band \textit{transverse} phasematching
function can be obtained by choosing a relatively large focusing
spot.  As was the case for the uncorrelated state, an ultrafast
pump is required with sufficient bandwidth to encompass the
spectral structure provided by the longitudinal and transverse
phasematching functions.  The condition thus imposed on the
crystal length $L$ and focused beam spot size $w_0$ can thus be
expressed as follows:
\begin{equation}
\frac{w_0}{L}\gg\frac{\sqrt{\gamma}(k_p'-k' \mbox{cos}\theta)}{k'
\mbox{sin}\theta}\label{E:poscorrcond}
\end{equation}
which should be compared with the condition guaranteeing spectral
uncorrelation [see Eq.~(\ref{E:Lw0})]. In addition
Eq.~(\ref{E:pumpreq}) needs to be fulfilled to ensure that the
pump has sufficient bandwidth.  Note that a strict frequency
correlated state would  have zero width along the
$\omega_s-\omega_i$ axis to ensure that the signal and idler
photon frequencies are identical. We can get a good approximation
to such a state by using a large focused beam spot size.
 Fig.~\ref{Fi:poscorr} shows a specific example of the synthesis a
frequency correlated state employing a $200 \mu\mbox{m}$-long
type-I BBO crystal with a cut angle of $\theta_{PM}=30.32^\circ$.
When pumping such a crystal with an ultrafast pulsed pump centered
at $\lambda_p=0.4~\mu\mbox{m}$, degenerate non-collinear PDC is
phase matched at propagation angles (within the crystal) of
$\theta_s=3^\circ$ and $\theta_i=-3^\circ$.  A focused beam spot
size of $w_0=1 \mbox{mm}$ ensures that the condition expressed in
Eq.~\ref{E:poscorrcond} is fulfilled yielding a frequency
correlated two-photon source over the bandwidth imposed by the
pump bandwidth and the crystal length as shown in
Fig.~\ref{Fi:poscorr}(D).

\section{Efficient generation of photon pairs}

In the preceding section, we simplified the analysis with the
assumption that the down-conversion output is collected from well
defined directions, which corresponds to fixing the values of the
transverse wave vectors in the joint two-photon probability
amplitude. The task of engineering the complete modal
characteristics of the down-converted photons has to include these
spatial degrees of freedom as well. A promising route towards
gaining control over the spatial structure of PDC is offered by
non-linear waveguides, in which the down-conversion process
generates photon pairs in the modes supported by the waveguide
structure. In this section, we review our recent experimental work
on this subject.

Nearly all PDC experiments to date have used as source a nonlinear
$\chi^{(2)}$ bulk crystal exploiting the principle of
bi-refringent phase-matching.   Observation of PDC in nonlinear
waveguides incorporating the principle of quasi phase matching
(QPM)  has been the subject of several recent experiments
\cite{BonfPrunAPL99,TanzDeRiElL01,SanaKawa,KonradAlfred,Gisin02}.
Two of them \cite{BonfPrunAPL99,TanzDeRiElL01} consisted of
detecting coincidences on the whole down-conversion signal divided
by a 50:50 beam splitter. The third experiment \cite{SanaKawa}
measured a Franson-type two-photon interference effect using an
unbalanced Michelson interferometer. Recently, Gisin \textit{et
al.} have used PDC generated in a periodically poled lithium
niobate (PPLN) waveguide to demonstrate time-bin entangled qubits
\cite{Gisin02}. In an experiment by our own group
\cite{KonradAlfred} we aimed at separating {\em all} the photon
pairs of interest into two different spatial paths with the help
of a spectrographic setup. Using this technique, we were able to
demonstrate the presence of strong correlations between photons of
different frequencies. We treated one path as a trigger while
collecting effectively all the conjugate photons in the second
signal path. The quantity of primary interest in our experiment is
thus the ratio of coincidences to single trigger counts, and the
setup presented below can be viewed as a scheme for generating
single ``heralded'' photons in the temporal slots defined by the
firing of the trigger detector.

\begin{figure} [htbp]
\vspace*{13pt}
\centerline{\psfig{file=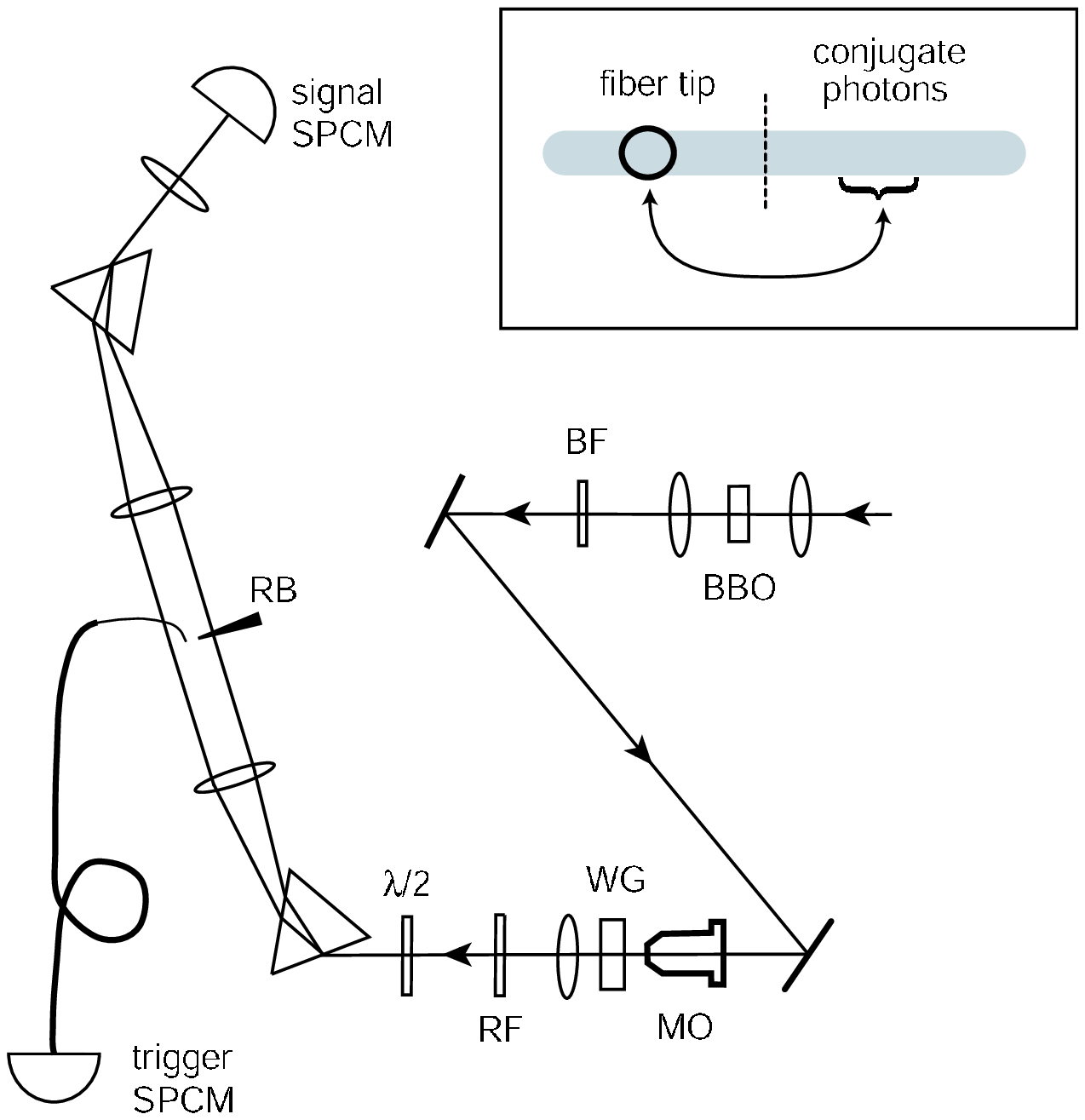, width=4.6cm}} 
\vspace*{13pt} \fcaption{\label{Fi:Setup}Experimental setup for
efficient generation of photon pairs using a KTP nonlinear
waveguide. }
\end{figure}

The experimental setup is depicted in Fig.~\ref{Fi:Setup}. The
output of a mode-locked Ti:Sapphire oscillator is first doubled in
a type-I BBO crystal to generate blue pulses which are focused on
the input face of a 1 mm long quasi-phase-matched KTiOPO$_4$ (KTP)
waveguide. The production and the characteristics of the sample
used in our experiment have been described elsewhere
\cite{Roel94}. The light power injected into the waveguide,
measured before the objective, is about 22~$\mu$W. The bandwidth
of the down-converted light for the parameters of our experiment
far exceeds 100~nm, and the down-converted photons are polarized
parallel to the pump field. A spectrographic setup is used to
separate the frequency anti-correlated signal and idler photons.
The maximum ratio of coincidence counts to singles observed in our
experiment is 18.5\%. This figure can be considered as the overall
detection efficiency of the signal photons, including all losses
from generation to detection.  Further details of the experiment
may be found in Ref.~\cite{KonradAlfred}.

The highest ratio of coincidences to singles reported in a PDC
experiment, exceeding $75\%$ at $702$~nm, was observed by Kwiat
{\em et al.} \cite{KwiatPRA93} It should be noted, however, that
this figure provided by the authors entailed the subtraction of
optical losses not due to the non-unit quantum efficiency of the
detectors.  The use of a long crystal ($10$~cm long KDP crystal)
introduced more stringent phase-matching conditions leading to
strong spatial correlations between the down-converted photons.

The utilization of QPM means that because the pump, signal and
idler photons can have the same polarization, the diagonal
elements of the $\chi^{(2)}$ tensor are accessible.  Because
diagonal elements are often higher in magnitude than non-diagonal
elements, this leads to higher production rates.  Furthermore, the
fact that photon generation is constrained to the modes of the
waveguide means that photons can be collected more efficiently as
compared to the case of a bulk crystal, where photons are emitted
in comparatively complicated spatial patterns. Our experiment
shows that a nonlinear waveguide can be used as a single
``heralded'' photon source with high brightness while using a
relatively short crystal and a relatively weak pump. Thus,
nonlinear waveguides have the very important advantage that PDC
can be generated economically as compared to bulk crystals.   To
illustrate this, Table~\ref{Ta:efficientPDC} shows the economy
figure of merit $R={R_s}/{LP}$ for several recent PDC experiments
where $R_s$ is the number of (single) photons detected per second,
$L$ the length of the crystal and $P$ the pump power used. As can
be seen, waveguide sources are associated with a value of $R$
several orders of magnitude higher that with bulk crystals.

\vspace*{4pt}   
\begin{table}[htbp]
\tcaption{Generating correlated photon pairs
efficiently.\label{Ta:efficientPDC}}
\centerline{\footnotesize\smalllineskip
\begin{tabular}{|r|r|r|r|r|r|r|}
\hline {REF.} & {CRYSTAL} & {$P [mW]$} & {$R_s [kHz]$} &
{$\frac{R_c}{R_s}$ [\%]} &
{$R [\frac{Hz}{m\cdot W}]$} \\
\hline \cite{KwiatPRA93} &
Type I 10cm KDP  & $.010$ & $65$  & 75 & $6.5 \times 10^7 $\\
\cite{Weinfurter2001} & Type II 2mm BBO & $465$ & $1250$  & 26& $2.7 \times 10^6$ \\
\cite{KonradAlfred} & 1mm KTP waveguide & $.022$ & 720  & 18.5& $3.3 \times 10^{10}$ \\
\hline
\end{tabular}}
\end{table}

\section{Application: Efficient linear optics quantum computation}

There have been recent proposals for building practical quantum
computers using passive linear optical systems
\cite{KLM,RalphPRA}. Ralph \textit{et al.} \cite{RalphPRA}
developed a simplification of the original Knill, Laflamme and
Milburn (KLM) \cite{KLM} proposal which envisions the nonlinear
sign-shift (NS) gate as a basic building block of quantum
computers. They show that a C-NOT gate can be constructed from two
such NS gates.  Here we focus on the photonic engineering
challenges that would be faced when implementing an NS gate using
PDC sources.

\begin{figure} [htbp]
\vspace*{13pt}
\centerline{\psfig{file=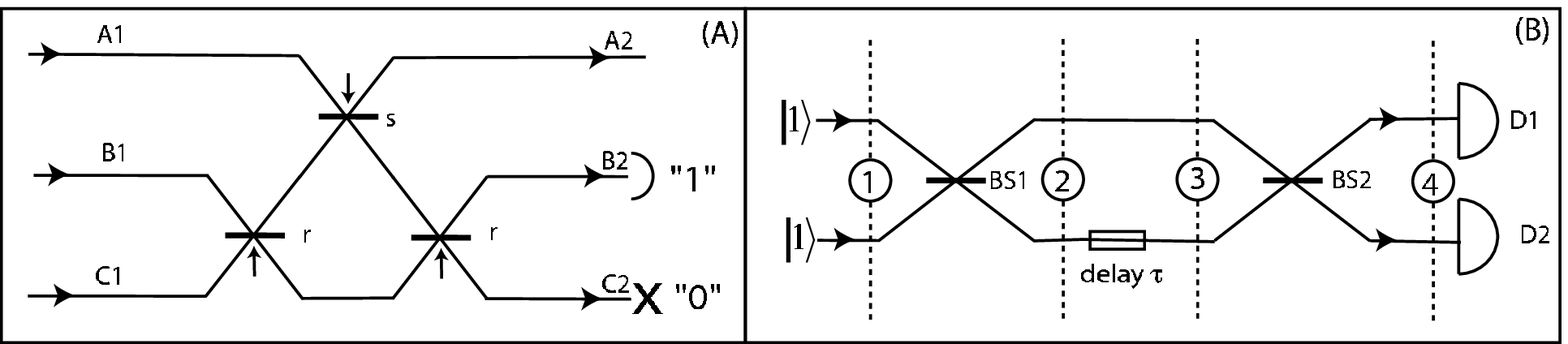, width=11.2cm}} 
\vspace*{13pt} \fcaption{\label{Fi:NSandMZ} (A) Non-linear sign
shift (NS) gate. (B) HOMI Mach-Zendher interferometer.}
\end{figure}

The NS gate is shown in Fig.~\ref{Fi:NSandMZ}(A) and its operation
is described in Ref.~\cite{RalphPRA}.  We assume the convention
that a $\pi$ phase shift occurs only upon reflection from one side
(indicated in Fig.~\ref{Fi:NSandMZ}(A) with a vertical arrow).
Suppose the inputs in the ancillary channels $B1$ and $C1$ are
$|1\rangle$ and $|0\rangle$ (where $|n\rangle$ is an $n$-occupancy
Fock state) respectively and that the signal input in channel $A1$
is occupied by the state $|\Psi_{\mbox{\scriptsize
IN}}\rangle=\alpha|0\rangle+\beta|1\rangle+\gamma|2\rangle$. It is
shown by Ralph \textit{et al.} that the firing pattern: ``1''
photon in ancillary output $B2$ and ``0'' photons in ancillary
output $C2$ implies that the $|2\rangle$ term has undergone a
$\pi$ phase shift, so that the signal output at $A2$ becomes:
$|\Psi_{\mbox{\scriptsize
OUT}}\rangle=\alpha|0\rangle+\beta|1\rangle-\gamma|2\rangle$. The
gate is non-deterministic in the sense that this firing pattern
occurs $25\%$ of the time for an optimal choice of beamsplitters
reflectivities $r$ and $s$ [see Fig.~\ref{Fi:NSandMZ}(A)].  Note
that the NS gate necessitates a detector which can discriminate
between 1 and 2 photons.   The proposal in Ref. \cite{Loop} may
help in realizing such discrimination.

\begin{figure} [htbp]
\vspace*{13pt}
\centerline{\psfig{file=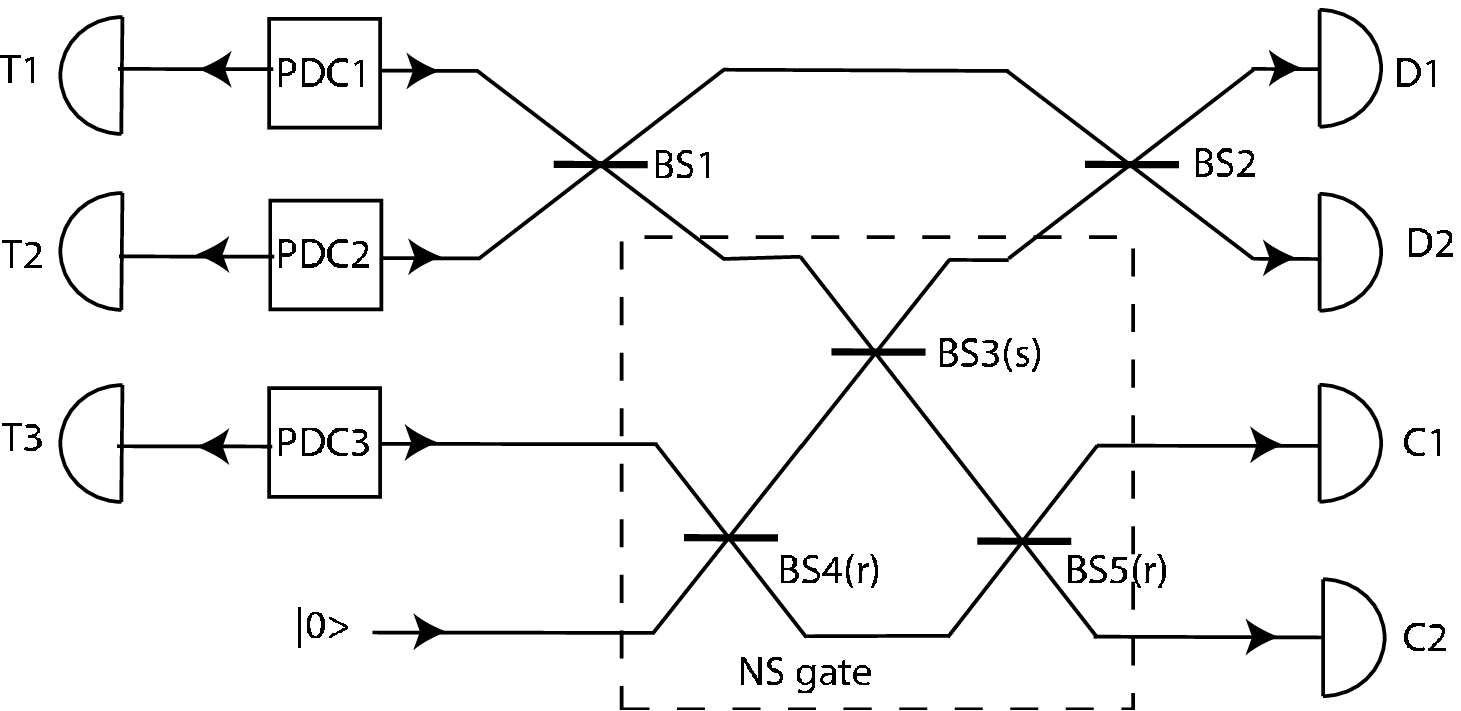, width=8.2cm}} 
\vspace*{13pt} \fcaption{\label{Fi:NS}An NS gate coupled to one
arm of a HOMI Mach Zehnder interferometer in an implementation
involving three PDC sources.}
\end{figure}

Let us turn our attention to the HOMI Mach-Zehnder interferometer
(HOMI-MZ) depicted in Fig.~\ref{Fi:NSandMZ}(B).   The two inputs
are each occupied by a single photon coming for instance from a
PDC crystal.  The quantum interference at $BS1$ turns the input
state $|\Psi_1\rangle=|1\rangle|1\rangle$  into
$|\Psi_2\rangle=(|0\rangle|2\rangle-|2\rangle|0\rangle)/\sqrt{2}$
[the subscripts label the interferometer stages as indicated in
Fig.~\ref{Fi:NSandMZ}(B)]. If there is no phase difference between
the two Mach Zehnder arms, $BS2$ simply reverts the state back to
the input i.e. $|\Psi_4\rangle=|1\rangle|1\rangle$, which leads to
a coincidence event across the two outputs.  However, if there is
a phase difference of $\pi$ between the two arms, the state before
$BS2$ becomes
$|\Psi_3\rangle=(|0\rangle|2\rangle+|2\rangle|0\rangle)/\sqrt{2}$,
which remains unaltered by the second beam splitter.  In this
case, there is no coincidence event across the two outputs.   Our
approach is to employ this sensitivity to the phase difference
between the two interferometer arms to test the action of the NS
gate.

We propose a practical implementation involving three PDC sources,
so that six photons are emitted simultaneously.  The proposed
apparatus is shown in Fig.~\ref{Fi:NS}: an NS gate is coupled to
one arm of a HOMI-MZ with three stacked PDC crystals acting as a
source of three photons conditioned on the firing of the triggers
$T1$, $T2$ and $T3$.  When the NS gate triggers (i.e. ``1'' photon
is recorded at detector $C1$ and none at detector $C2$) the lower
interferometer arm undergoes a $\pi$ phase shift, which means that
the photons emerging at $D1$ and $D2$ are both in the same (but
undetermined) channel so that no $D1,D2$ coincidence event occurs.
Therefore, we may be convinced that the simultaneous firing of
detectors $T1,T2,T3,D1,D2,C1$ is an impossibility.  A different
way of saying this is that a $T1,T2,T3,D1,D2,C1$ coincidence event
implies that the NS gate did not fire.   We use this fact to test
the performance of the NS gate.  If the probability of obtaining
such a six-fold coincidence event is non-zero, the correlation
between the $C1,C2$ and $D1,D2$ firing patterns is imperfect.

\begin{figure} [htbp]
\vspace*{13pt}
\centerline{\psfig{file=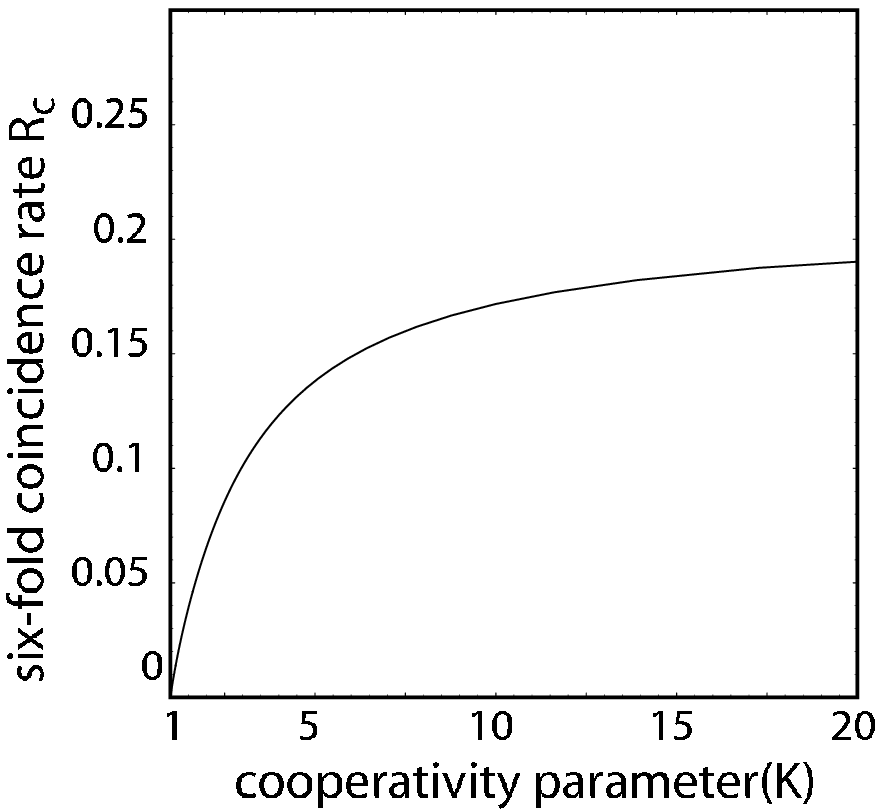, width=6.2cm}} 
\vspace*{13pt} \fcaption{\label{Fi:RcK} Relationship between the
six fold $T1 T2 T3 D1 D2 C1$ coincidence rate and the
cooperativity parameter.  In the ideal case where the firing
patterns at C1 and C2 are perfectly correlated to those at D1 and
D2, the coincidence rate must vanish.   Note that for $K>1$, i.e.
for a source which exhibits spectral correlations the NS gate does
not operate ideally.}
\end{figure}

The system described here involves multiple-source quantum
interferometry of the kind described in Sec.~\ref{Sec:eng}. We
would therefore expect the arguments about indistinguishability in
the form of spectral correlation to apply so that the system
should work perfectly only in the case where a single Schmidt mode
is active (i.e. $K=1$) in each of the three PDC sources. In order
to model this apparatus, we assume that the three PDC sources are
described by a simplified type-I joint spectral amplitude as given
by Eq.~(\ref{E:ModelSource}).   The six-fold $T1,T2,T3,D1,D2,C1$
coincidence rate is given by $R_c=\int...\int
dt_1...dt_6|\hat{E}_1^{(-)}(t_1)...\hat{E}_6^{(-)}(t_6)|\Psi\rangle|^2$
where  $\hat{E}_j^{(-)}(t)$ with $j=1-6$ are the electric field
operators at each of the six output channels and take into account
the various beam-splitter transformations.  The quantum state is
assumed to be given by: $|\Psi\rangle=\int...\int
d\omega_1...d\omega_6
f(\omega_1,\omega_2)f(\omega_3,\omega_4)f(\omega_5,\omega_6)\hat{a}_1^\dag...\hat{a}_6^\dag|\mbox{vac}\rangle$
where each function $f(\omega_i,\omega_j)$ is given according to
Eq.~(\ref{E:ModelSource}). We further assume that the beamsplitter
reflectivities are given by\cite{RalphPRA}: $r=(4-2\sqrt{2})^{-1}$
and  $s=(\sqrt{2}-1)^2$.  Such a calculation yields the result
expressed in the plot of coincidence rate vs. cooperativity
parameter in Fig.~\ref{Fi:RcK}.  Note that a zero coincidence rate
is obtained, as expected, only in the case where there is no
spectral correlation.

\section{Conclusions}
Designing PDC sources with characteristics optimized for quantum
information processing applications presents a highly nontrivial
task. We have shown in this paper that a careful analysis of the
spatio-temporal characteristics of the down-conversion process
yields possible routes towards practical solutions to this
problem. These results, combined with the possibility of
engineering non-linear two-photon states by using waveguide
structures, open the prospects for the next generation of
down-conversion sources with the characteristics tailored
specifically for quantum information experiments. We have
illustrated the importance of source engineering with a discussion
of optical implementations of simple quantum logic circuits such
as the nonlinear phase shift gate.

\nonumsection{Acknowledgements} \noindent We have become aware
since preparing this manuscript that W.P.Grice and Y. Kim have
arrived at similar conclusions to the ones presented in
Sec.~\ref{Sec:criteriapolent}\cite{warrenLANL}.  The authors
acknowledge insightful discussions with W.P. Grice, M. Teich, A.
Sergienko and M. Raymer.   This work was partially supported by
ARO-administered MURI grant number DAAG-19-99-1-0125 and by the
National Science Foundation (NSF).

\nonumsection{References}

\appendix

\noindent While carrying out a Schmidt decomposition analytically
for an arbitrary two-photon state is non-trivial, for certain
classes of states it may be achieved.   We will find the following
identity useful:

\begin{equation}
\begin{split}
\mbox{exp}\left[-\frac{1+\mu^2}{2(1-\mu^2)}(\alpha_1^2 x_1^2+\alpha_2^2 x_2^2)+\frac{2  \alpha_1 \alpha_2 \mu x_1 x_2}{1-\mu^2}\right]=\\
\sqrt{1-\mu^2}\sum\limits_{n=0}^{\infty}\mu^n u_n(\alpha_1
x_1)u_n(\alpha_2 x_2)
\end{split}
\label{E:identity}
\end{equation}
valid for $\mu\ne 0$ and where $u_n(x)=(2^n
n!)^{-\frac{1}{2}}H_n(x)\mbox{exp}(-x^2/2)$, $H_n(x)$ being the
$n^{th}$ order Hermite polynomial.   Any two photon state
characterized by a spectral amplitude $f(\omega_s,\omega_i)$ which
can be written in the form of the left-hand-side of
Eq.~(\ref{E:identity}) can be decomposed into its Schmidt
functions with the help of this identity.   Comparing
Eq.~(\ref{E:identity}) with Eq.~(\ref{E:schmidt}), we can write
down an expression for the eigenstates $\lambda_n$:

\begin{equation}
\lambda_n=(1-\mu^2)\mu^{2n} \label{E:eigenvalues}
\end{equation}

By substituting Eq.~(\ref{E:eigenvalues}) into Eq.~(\ref{E:K}), we
obtain an expression for the cooperativity parameter:

\begin{equation}
K=\frac{1+\mu^2}{1-\mu^2} \label{E:Kanalytical}
\end{equation}

The two-photon state characterized by the idealized type-I
two-photon state introduced in Sec.~\ref{Sec:criteria} [see
Eq.~(\ref{E:ModelSource})] belongs to the class of sates
decomposable through Eq.~(\ref{E:identity}). Carrying out the
decomposition we find that:

\begin{equation}
\mu=1+\left(\frac{\sigma}{\sigma_F}\right)^2-\sqrt{2
\left(\frac{\sigma}{\sigma_F}\right)^2+\left(\frac{\sigma}{\sigma_F}\right)^4}
\label{E:mumodelsource}
\end{equation}

We can now obtain an expression for the cooperativity parameter
for the type-I idealized source in terms of the source parameters
$\sigma$ and $\sigma_F$ by substituting
Eq.~(\ref{E:mumodelsource}) into Eq.~(\ref{E:Kanalytical}).  Such
an expression was used to generate the plots in
Fig.~\ref{Fi:visKvisR}.

\end{document}